\documentclass[aps,prfluids,11pt]{revtex4-2}
\usepackage{amsmath}
\usepackage{amssymb}
\usepackage{mathcomp}           
\usepackage{textcomp}           
\usepackage{float}              
\usepackage{hyperref}           
\usepackage{setspace}           
\usepackage{siunitx}            
\usepackage{soul}               
\usepackage{xcolor}             
\usepackage{graphicx}           
\usepackage{dcolumn}            
\usepackage{bm}                 
\usepackage{silence}            
\usepackage{placeins}           
\usepackage{enumitem}           

\newcommand{\hz}{ \hspace{2pt}} 

\vbadness=\maxdimen     

\newcommand{\degC}{\ensuremath{\tccentigrade}}

\begin{document}

\title{Hidden in plain sight: How evaporation impacts the pendant drop method}

\author{Pim J. Dekker}
\author{Christian Diddens}
\author{Marjolein N. van der Linden}
\altaffiliation[Also at ]{Canon Production Printing Netherlands B.V., P.O. Box 101, 5900 MA Venlo, The Netherlands}
\author{Detlef Lohse}
\altaffiliation[Also at ]{Max Planck Institute for Dynamics and Self-Organization, Am Fassberg 17, 37077 G\"ottingen, Germany}
\affiliation{Physics of Fluids group, Department of Science and Technology, Max Planck Center for Complex Fluid Dynamics and J. M. Burgers Centre for Fluid Dynamics, University of Twente, P.O. Box 217, 7500 AE Enschede, The Netherlands}

\date{\today}

\begin{abstract}
    The surface tension of a liquid, which drives most free surface flows at small scales, is often measured with the pendant drop method due to its simplicity and reliability. When the drop is suspended in air, controlling the ambient temperature and humidity is usually an afterthought, resulting in evaporation of the drop during the measurement. Here, we investigate the effect of evaporation on the measured surface tension using experiments and numerical simulations. In the experiments, we measured the evolution of the droplet temperature, which can drastically reduce by ($\Delta T \approx 10 \degC$) due to evaporative cooling, and thereby altering the measured surface tension by more than 1 mN/m. This finding can be reproduced by numerical simulations, which additionally allows for controlled investigations of the individual influences of further effects on the pendant drop method, namely shape deformations by evaporation-driven flows in the gas-phase and in the liquid-phase including the resulting Marangoni flow. We provide a simple passive method to control the relative humidity without requiring additional instrumentation. Our findings are particularly pertinent to Marangoni flows which are driven by surface tension gradients, and which are consequently highly sensitive to measurement inaccuracies. We apply our method with different aqueous mixtures of glycerol and various diols. Our results and insights have implications for various applications, ranging from inkjet printing to agricultural sprays. Finally, we have meticulously documented our setup and procedure for future reference.
\end{abstract}

\maketitle


\section{Introduction}
The surface tension of a liquid-air interface plays a central role in the majority of free surface flows at small scales. Examples include: jetting, pinch-off, drop impact, spreading, coalescence, and evaporation \cite{vanderbos2014,lohse2022,herman2018,bonn2009,eggers2024,wang2022,wilson2023}. Moreover, surface tension gradients can have a profound impact on the flow in a system. Particularly for evaporating sessile drops, where surface tension gradients can arise due to the non-uniform evaporative flux which results in temperature gradients due to evaporative cooling, or compositional gradients due to selective evaporation \cite{gelderblom2022}. Even small differences in surface tension ($< 0.5 \ \mathrm{mN/m}$) can completely alter the flow in the drop, such as the flow magnitude, direction of the flow, and even cause instabilities \cite{seyfert2022,baumgartner2022,li2018,diddens2024}. Therefore accurate and precise surface tension measurements are crucial for understanding of free surface flow.

Various methods have been developed to measure surface tension \cite{adamson1997,ebnesajjad2011}, such as: the capillary rise method, the du No\"uy ring method, the pendant drop method, the Wilhelmy plate method, maximum bubble pressure method, and the rotating drop method. Each method has its own advantages and limitation, so careful consideration of the measurement requirements is needed \cite{ebnesajjad2011}. The pendant drop stands out from these techniques as, other than a camera and lens (even just a smartphone camera \cite{goy2017}), no additional instrumentation is required, making it very popular to determine the surface tension when specialised equipment such as a force tensiometer is not available. 

Although it is obvious that a pendant drop evaporates when measuring the surface tension of mixture containing volatile components, most notably water, controlling evaporation is usually an afterthought. However, it remains unclear to what extend evaporation plays a role when measuring the surface tension using the pendant drop method. In this work we experimentally and numerically investigate the role of evaporation on the accuracy of the pendant drop method.

This work also serves as the companion to the paper \cite{diddens2024}, which shows through numerical simulations that an instability occurs for an evaporating water/1,2-hexanediol drop because the surface tension of water/1,2-hexanediol is non-monotonic, which is a testament to the importance of surface tension on the evaporation dynamics. This work experimentally confirms that the surface tension of water/1,2-hexanediol is indeed non-monotonic using the pendant drop method and the dy No\"uy ring method. Additionally, we measure the surface tension for various aqueous mixtures n-diols in search of an rationale for the unusual case of 1,2-hexanediol. Although we find a clear trend with increasing chain length, the exact mechanism that makes 1,2-hexanediol non-monotonic remains elusive. 

The paper is structured as follows: In section \ref{sec:method}, we introduce the different measurement principles for surface tension and discuss the experimental methodology, including the experimental setup, procedure, image processing, and error analysis. Although the pendant drop method is relatively straightforward in principle, special attention is required to attain sufficiently accurate and precise measurements. In section \ref{sec:evaporation}, we experimentally and numerically show that a water drop cools down significantly ($\sim 10 \mathrm{^\circ C}$) at low relative humidity due to evaporative cooling. Additionally, we show that both natural convection around the drop and thermal Marangoni flow in the drop have a negligible effect on the shape of the drop and therefore negligible effect on the measured surface tension. Finally, in section \ref{sec:surface_tension}, we experimentally determine the surface tension of water/1,2-hexanediol with different methods and compare to other n-diols.



\section{Methods} \label{sec:method}

\subsection{Surface tension measurement principles}

Next to the pendant drop method, we will also discuss two other methods. First, the capillary rise method, since we will compare the surface tension of water/1,2-hexanediol to a previous measurement with this method \cite{romero2007fpe}. Second, the du No\"uy ring method, since this method is one of the most widely used methods next to the pendant drop method. 

\begin{figure}[tbp]
    \centering
    \includegraphics[width=1\textwidth]{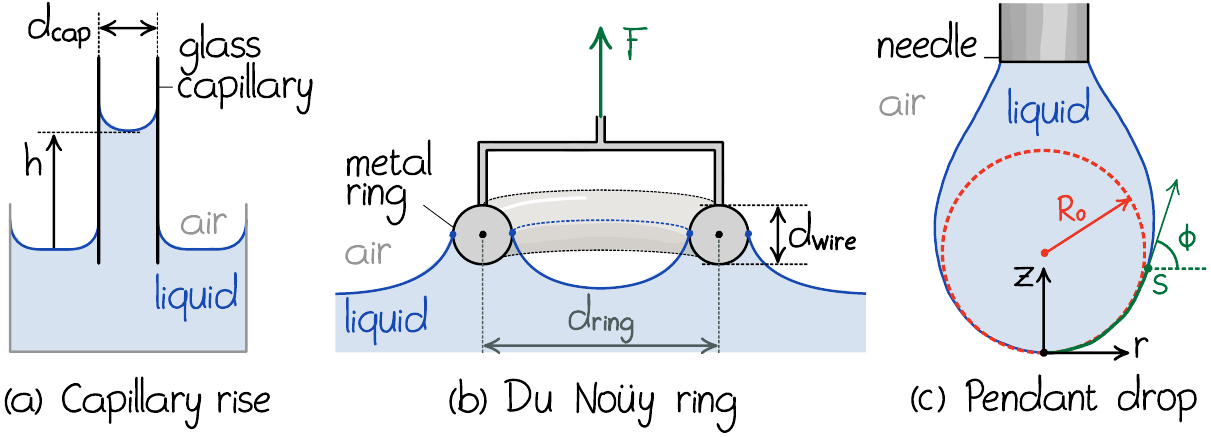}
    \caption{Overview of different methods used to measure surface tension. (a) Capillary rise method. Here, $h$ is determined by the balance of surface tension and gravity. (b) Ring tensiometer, which directly measures the surface tension, although it must be compensated for gravity. (c) Pendant drop method. Here, the shape of the drop is determined by a balance between surface tension and gravity.}
    \label{fig:methods}
\end{figure}

The capillary rise method measures the surface tension by placing a glass tube with a small inner diameter vertically with one end in the sample liquid \cite{rayleigh1916,cao2023}. When the liquid is completely wetting, a meniscus will form inside the capillary and pull up the liquid, as is shown schematically in figure \ref{fig:methods}(a). The final height, $h$, is a balance between the surface tension, $\gamma$, and gravity and is given by the following equation \cite{rayleigh1916}
\begin{equation}
    \gamma=\frac{\Delta\rho g r_\mathrm{cap}}{2}\left(h+\frac{r_\mathrm{cap}}{3}-0.1288\frac{ r_\mathrm{cap}^{2}}{h}+0.1312\frac{ r_\mathrm{cap}^{3}}{h^{2}}\right).
    \label{eq:gamma_cap}
\end{equation}
Here, $\Delta\rho$ is the difference in density between the liquid and the surrounding air, $g$ is the gravitational acceleration, and $r_\mathrm{cap}$ is the inner radius of the capillary. The two last terms in eq. \ref{eq:gamma_cap} are the first two leading order corrections for the non-sphericity of the meniscus shape. When the capillary radius is small with respect to the capillary length, $l_\mathrm{c} = (\gamma/\Delta\rho g)^{1/2}$, the correction terms will be negligible. For liquids which are not completely wetting, eq. \ref{eq:gamma_cap} must be corrected for the contact angle and the factors taking the meniscus shape into account change \cite{adamson1997}.

The du No\"uy ring methods determines the surface tension by pulling a thin ring out of the liquid and measuring the maximum force, $F$, which coincides with the surface tension pulling vertically downwards on the perimeter of the ring \cite{dunouy1919}. The surface tension is then given by
\begin{equation}
    \gamma=\frac{F}{4 \pi r_\mathrm{ring}} f.
\end{equation}
However, there is an asymmetry between the liquid-vapor interface surrounding the ring, where the curvature of the interface is zero, and the interface enclosed by the ring, where the curvature is not perfectly zero. This changes the measured surface tension significantly ($\sim 25\%$), even if the radius of the ring, $r_\mathrm{ring}$, is much larger than the capillary length. A correction factor $f$ is introduced to correct for the asymmetry, which can be estimated by different models \cite{harkins1930,zuidema1941,huh1975}. In general, $f$, depends on the geometry of the ring, the height of the ring above the liquid surface, and the capillary length \cite{adamson1997}. 

The pendant drop methods determines the surface tension by measuring the shape of a pendant drop, which is determined by surface tension and gravity \cite{stauffer1965,rotenberg1983,berry2015}. Assuming that the surface tension is constant and the drop is in equilibrium, the pressure across the liquid-vapor interface must be constant, $\Delta p_0 = \Delta p_g + \Delta p_\gamma$, and is given by
\begin{align}
    \frac{2\gamma}{R_0} & = \Delta\rho g z  + 2 \kappa \gamma.
\end{align}
Here, we defined $z=0$ as the bottom of the drop, $R_0$ is the radius of curvature at the bottom of the drop and $\kappa$ is the curvature. Making this equation dimensionless using $R_0$, we obtain
\begin{align}
2  = \mathrm{Bo}\, \tilde{z}+ 2 \tilde{\kappa}, \qquad \mathrm{Bo} = \frac{\Delta\rho g R_0^2}{\gamma}.
\label{eq:pen_dimles}
\end{align}
Here, Bo is the Bond number which is the ratio of gravity over surface tension. To be able to compute the shape of the drop, we parametrize eq. \ref{eq:pen_dimles} in terms of the arc length, $s$, and the angle w.r.t. the horizontal, $\phi$, which gives a set of ordinary differential equations:
\begin{align}
    \frac{d\phi}{d\tilde{s}} = 2 - \mathrm{Bo} \,\tilde{z} -  \frac{\sin{\phi}}{\tilde{r}}, \qquad  \frac{d\tilde{r}}{d\tilde{s}} = \cos{\phi}, \qquad  \frac{d\tilde{z}}{d\tilde{s}} = \sin{\phi},
    \label{eq:fit_YL_eq}
\end{align}
which can be numerically integrated using the following boundary conditions:
\begin{align}
    \phi(\tilde{s}=0) = 0, \qquad \tilde{r}(\tilde{s}=0) = \varepsilon, \qquad  \tilde{z}(\tilde{s}=0) = 0.
    \label{eq:fit_YL_BC}
\end{align}
Here, $\varepsilon$ is a very small number to avoid the singularity at $\tilde{r} = 0$ and is chosen to be sufficiently small such that the solution is independent of $\varepsilon$. The surface tension is calculated by fitting $R_0$, $z_0$, $r_0$, and Bo to the drop shape. For the pendant drop method to be accurate, the effect of gravity must be clear and the drop shape must be sufficiently non-spherical \cite{berry2015}.

\FloatBarrier

\subsection{Experimental procedure and analysis of the pendant drop method}

\begin{figure}[tb]
    \centering
    \includegraphics[width=0.6\textwidth]{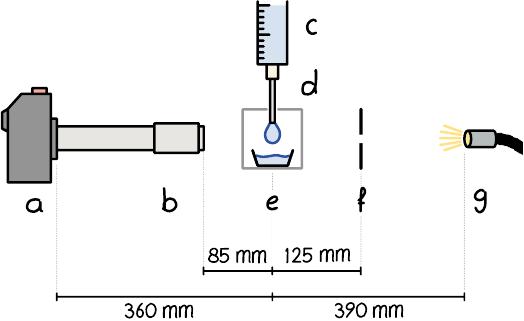}
    \caption{Schematic of the setup used for the pendant drop measurements. a: Camera (D850, Nikon), b: Long distance microscope (12$\times$ Zoom, Navitar), c: Lubricant-free syringe (Injekt Luer Solo, B. Braun), d: Needle (14 gauge, Metcal), e: Glass cuvette ($50 \times 50 \times 50$ mm, Hellma) with reservoir, f: Aperture ($d = 7.0$ mm), g: Light source (KL2500, Schott).}
    \label{fig:setup}
\end{figure}

A schematic of the setup for the pendant drop measurements is shown in figure \ref{fig:setup}. Both the syringe and the needle are lubricant free, i.e. without silicon. The light was placed at a large distance from the drop to ensure the rays are as parallel as possible without needing additional optics. A pinhole ensures that reflections are minimised. The syringe is manually actuated using a high precision translation stage (PT1/M, Thorlabs), which means no tubing is required that otherwise might introduce unnecessary contamination.

All liquids except water were supplied by Sigma Aldrich and were used without further purification (purities: 1,2-butanediol $\geq97.0\%$; 1,2-pentanediol $\geq96.0\%$; 1,2-hexanediol $\geq98.0\%$; 1,5-pentanediol $\geq97.0\%$; glycerol $\geq99.5\%$ ). Ultra pure water was obtained using a Milli-Q IQ 7000. Samples were stored in glass vials to minimise any effect of plastic leaching. Before use, all equipment that contact the sample solutions are thoroughly cleaned by rinsing multiple times with acetone, ethanol, and water. A pendulum was used to ensure that the needle was perfectly vertical. Before each measurement, some of the sample liquid was flushed through the needle to ensure that the next drop would be as close as possible to the concentration of the sample liquid and to the surrounding temperature of the lab. The image was taken a few seconds after the drop was produced to ensure all vibrations had been subsided. For the time series measurements, the images were continuously taken at a fixed interval.

Evaporation is controlled passively by placing a reservoir containing the sample liquid in the chamber. This will ensure that the relative humidity in the chamber will equilibrate, ceasing all evaporation or condensation. Not only is this method very straightforward to implement, but also has the advantage that no airflow is induced in the chamber, which is inevitable with an active humidity control method. Additionally, while an active method needs constant adjustments depending on temperature in the chamber or composition of the sample liquid, this is not the case for a passive method. To achieve very low humidities, anhydrous CaCl$_2$ beads are placed in the bottom of the chamber, which absorb nearly all water vapor, resulting in a constant low relative humidity. The temperature and relative humidity in the chamber were measured using a digital sensor (BME280, Bosch Sensortec) connected to a microcontroller (Feather M4 Express, Adafruit) that recorded the data to a PC \footnote{Source code humidity and temperature sensor (Python): \url{https://github.com/Dennis-van-Gils/project-Humidistat}}.
\begin{figure}[tb]
    \centering
    \includegraphics[width=0.9\textwidth]{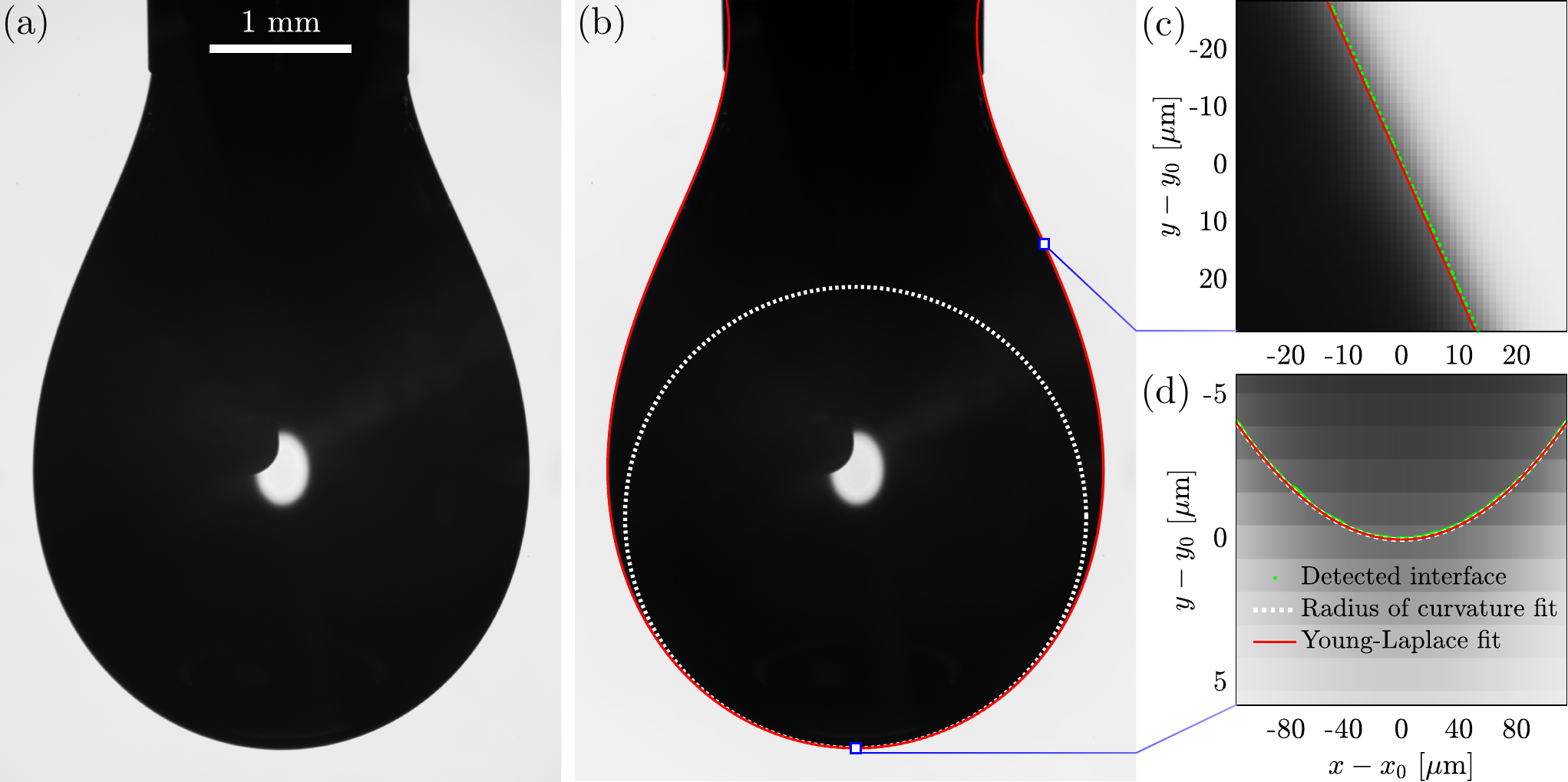}
    \caption{Experimental image of a pendant drop of water optimised for contrast. (a) The raw image. (b) The image corrected for noise and illumination. The detected interface (green dots) is obscured by the fit of the Young-Laplace equation (red line). The white discontinuous line is the fit of the radius of curvature. The scale is identical to (a). (c) Close up of the drop interface as indicated by the blue line and blue marker in (b). (d) Close up of the drop interface at the bottom. Note that the horizontal and vertical scale in (d) are different.}
    \label{fig:drop_and_fit}
\end{figure}

The drop is imaged with a camera and long distance microscope. A typical image is shown figure \ref{fig:drop_and_fit}(a). First we apply a Gaussian blur with a standard deviation of two pixels to remove noise. This step does not reduce the accuracy in the final detected edge since the features of the drop are much larger than two pixels. Additionally, the image is not perfectly sharp anyway since the camera resolution ($\sim 1 \ \mathrm{\tcmu m}$) is much smaller than the optical resolution ($\sim 8 \ \mathrm{\tcmu m}$). Next, the intensities are normalised using the background calibration, the details of which are in the appendix (figure \ref{fig:light}). The lighting-corrected and noise reduced image is shown in figure \ref{fig:drop_and_fit}(b).

To obtain the drop interface from the image we use a threshold with subpixel accuracy by interpolating the intensity with the threshold for each column and row of pixels. The bright spot of the light in the drop center is ignored when determining the drop edge. The region close to the needle (0.1 mm) is also ignored. Although more sophisticated edge detection algorithms exist, in this case their added value, beyond longer computational time, is limited since the edge is very clean and consistent throughout the image. In fact, using a threshold approach gives precise and intuitive control of the location of the detected edge, which is important as we will later show. Next, the resolution calibration is applied to convert coordinates of the detected edge from pixels to meters, where we take special care to account for lens distortion. Finally, corrections were applied for the angle of the camera by rotating the detected interface. Details of the resolution calibration (figure \ref{fig:resolution}) and the angle calibration (figure \ref{fig:angle}) are provided in the appendix.

We fit the Young-Laplace (YL) equation to the so-obtained drop interface. This is done in two steps. First, we fit a circle to the bottom $1.5\%$ of the drop interface, resulting in the radius of curvature, $R_0$, at the bottom of the drop, as well as the center of the circle, which we both use to make the coordinates of the interface dimensionless. Second, we fit the YL equation to each half of the drop by varying the Bo number, resulting in two Bo numbers, $\mathrm{Bo_-}$ for the left half and $\mathrm{Bo_+}$ for the right half. The final Bo number is the average between the two. Both the circle fit and the YL fit are shown in figure \ref{fig:drop_and_fit} and show excellent agreement with the detected interface points. The fitting procedure and the residues are shown in detail in the appendix in figure \ref{fig:exp_R0_fit} and \ref{fig:exp_Bo_fit}.

\begin{figure}[tb]
    \centering
    \includegraphics[width=0.85\textwidth]{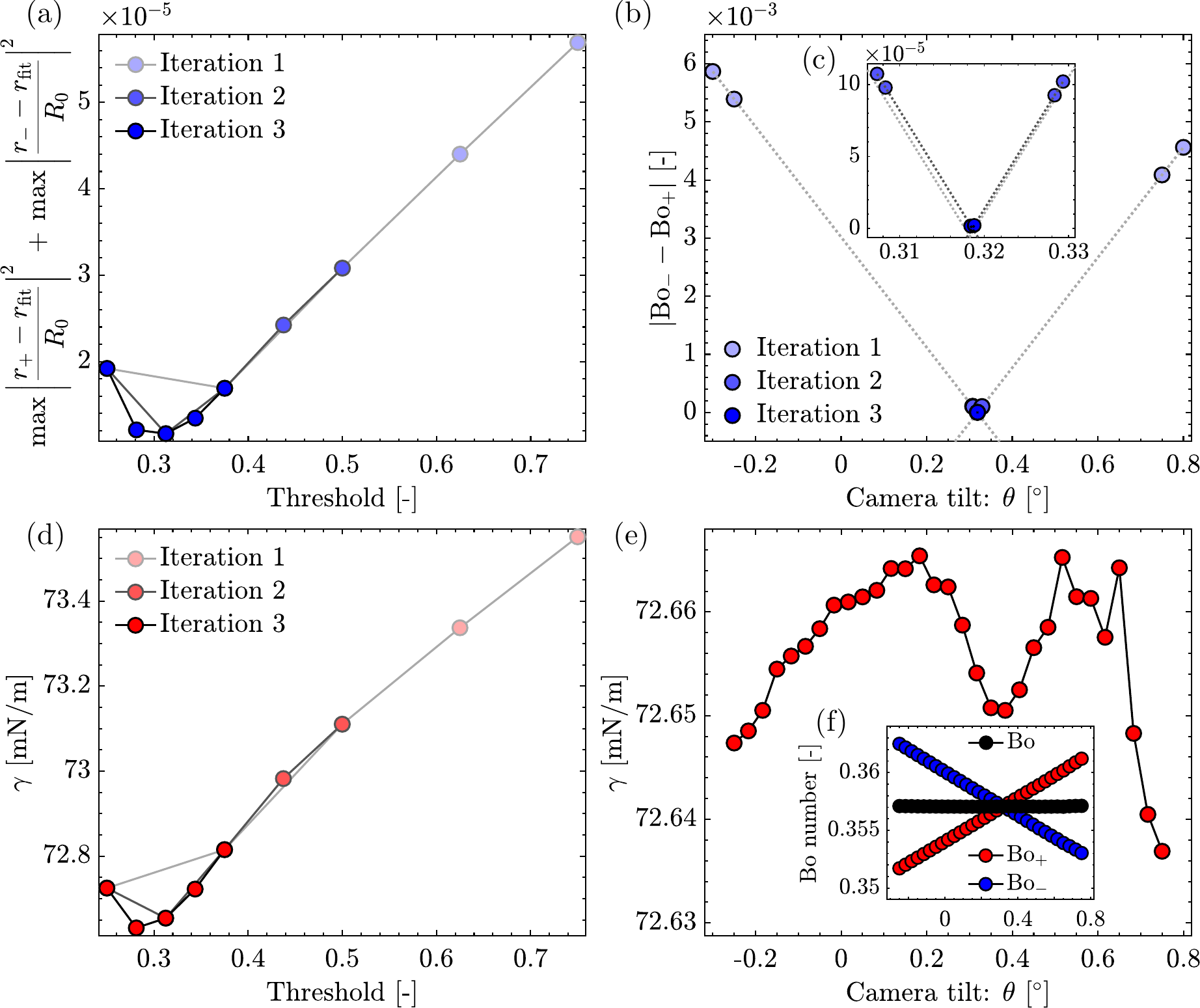}
    \caption{(a) The optimal threshold value is determined by finding the minimal deviation between the YL-fit and the detected interface. By iterating, the computational time is minimised. (d) Shows how the measured surface tension depends strongly on the used threshold value, with almost 1 mN/m, and that the minimum in surface tension coincides with the minimum in deviation from the fit. (b) The camera angle is determined by minimising the difference between the Bo number for the left and right side of the drop. Since the relation is almost linear a fit can be used to dramatically reduces the range between iterations. (c) Shows a close up of the second iteration. (e) Shows that the measured surface tension does not depend strongly on the applied camera angle correction, with a maximum difference of only $\pm 0.02$ mN/m. (f) Shows that both the Bo number for the left and for the right side of the drop depend strongly on the applied camera angle correction. However, this dependence cancels out in the average Bo number, $\mathrm{Bo} = (\mathrm{Bo}_- + \mathrm{Bo}_+ /)2$.}
    \label{fig:exp_thres_theta}
\end{figure}

However, upon closer inspection of figure \ref{fig:drop_and_fit}(c) and (d), we find the that interface is a smooth gradient with a width of approximately 10 pixels ($\approx 10 \ \mathrm{\tcmu m}$). This is far larger than the deviation between the detected interface and the YL fit ($< 1 \ \mathrm{\tcmu m}$). Therefore, the choice of threshold value that is used to determine the interface will greatly affect the detected shape of the drop and consequently also the fit of the YL equation. Figure \ref{fig:exp_thres_theta}(d) shows that the apparent surface tension changes significantly, almost 1 mN/m, depending on the threshold value.

By minimising the error between the detected interface and the YL fit the most accurate threshold value can be determined systematically. Figure \ref{fig:exp_thres_theta}(a) shows the error in the YL fit, which is defined as the sum of the maximum deviation between the fit and the interface squared of both sides of the drop. The minimum in the error coincides with the minimum in surface tension, which was generally the case. For each consecutive series of images, the optimal threshold value was determined only once and then applied to all images. Despite normalising the intensities before determining the optimal threshold value, we found that the threshold value ranged from 0.3 to 0.55 between different measurement series. It is known that the threshold value can change the measured surface tension \cite{kalantarian2013}. Alternatively, a gradient based method might be used, which detects the edge using the maximum gradient or the inflection point in the intensity. However, such methods do not allow for fine tuning of the drop shape with a parameter similar to the threshold value \cite{vanderbos2014}.

Additionally, we also investigated the effect of the camera angle on the measured surface tension, which is shown in figure \ref{fig:exp_thres_theta}(e). The surface tension varies only slightly ($\sim$ 0.01 mN/m) with the camera angle. However, the left and right Bo numbers change significantly with the camera angle, as shown in the inset (f). By minimising the difference between the left and right Bo number, as shown in figure (b), we can systematically find the optimal camera angle, which agrees well with the externally calibrated camera angle (figure \ref{fig:angle}).

\begin{table}[tb]
    \caption{\label{tab:errors}
    Typical values and uncertainties of the relevant physical quantities and the resulting uncertainties on the final surface tension $\gamma$.
    }
    \begin{ruledtabular}
    \begin{tabular}{llll}
    \textbf{Physical quantity} & $\qquad$ \textbf{Value} & $\quad$\textbf{Uncertainty} & \textbf{Resulting uncertainty on $\gamma$} \\
    \colrule
    Resolution          & 1.0048 \hz [$\mathrm{\tcmu m\,px^{-1}}$] & $1.0 \cdot 10^{-5}$  [$\mathrm{\tcmu m\,px^{-1}}$]    & $\ \qquad $ 0.142 [$\mathrm{mN\,m^{-1}}$] \\
    Radius of curvature (fit) & 1.6301 \hz [mm]                    & $3.6 \cdot 10^{-5}$  [mm]                    & $\ \qquad $ 0.032 [$\mathrm{mN\,m^{-1}}$] \\
    Density             & 998.0  \hz \hz   [$\mathrm{kg\,m^{-3}}$] & $5.0 \cdot 10^{-1}$  [$\mathrm{kg\,m^{-3}}$] & $\ \qquad $ 0.036 [$\mathrm{mN\,m^{-1}}$] \\
    Bo number           & 0.35585          [\ensuremath{-}]        & $2.3 \cdot 10^{-4}$  [\ensuremath{-}]        & $\ \qquad $ 0.049 [$\mathrm{mN\,m^{-1}}$] \\
    Threshold           & 0.34 \hz \hz \hz [\ensuremath{-}]        & $2.0 \cdot 10^{-2}$  [\ensuremath{-}]        & $\ \qquad $ 0.032 [$\mathrm{mN\,m^{-1}}$] \\
    Camera angle        & 0.35936          [\ensuremath{^\circ}]   & $5.0 \cdot 10^{-2}$  [\ensuremath{^\circ}]   & $\ \qquad $ 0.009 [$\mathrm{mN\,m^{-1}}$] \\
    \end{tabular}
    \end{ruledtabular}
\end{table}

The uncertainty in the measured surface tension can be estimated by considering the uncertainties in the various quantities that are used to calculate $\gamma$.
\begin{equation}
    \gamma = \frac{g \Delta \rho R_0^2}{\mathrm{Bo}}
    \label{eq:gamma_bo}
\end{equation}
Typical values and the uncertainty for each physical quantity, and the resulting uncertainty on the measured surface tension are provided in table \ref{tab:errors}. The uncertainty in the resolution is based on the difference between the quartic (i.e. biquadratic) fit and the measured local resolution, see appendix. The uncertainty in the fit of the radius of curvature is calculated as the standard deviation of the difference between the fitted $R_0$ and the detected interface, see appendix. The uncertainty in the density is given by the device and observed fluctuations during the density measurement. The error in the YL fit of the Bo number is calculated as follows. First the sensitivity, $\partial r/\partial \mathrm{Bo}$, of the radius close to the needle to a small change in the Bo is determined at the final Bo number. We evaluate the sensitivity close to the needle since here the drop shape is the most sensitive to the Bo number. Then, using the root-mean-square (RMS) of the difference between the interface and the YL fit, we calculate the uncertainty in Bo as follows: 
\begin{equation}
    \mathrm{Bo}_\mathrm{uncertainty} = \bigg[ \frac{\operatorname{RMS}(r_- - r_\mathrm{fit}) + \operatorname{RMS}(r_+ - r_\mathrm{fit})}{2} \bigg]\left[  \frac{\partial r}{\partial \mathrm{Bo}} \right]^{-1}
\end{equation}
Although we already estimated the uncertainty in all physical quantities, we must also consider the sensitivity of the surface tension to the threshold value and the camera angle. This is done by varying the threshold value and the camera angle in the range of the each uncertainty and taking the minimum and maximum values of the resulting calculated surface tensions. The values in table \ref{tab:errors} differ for each measurement, but give a fair representation of the uncertainties. In almost all cases the uncertainty in the resolution (and therefore $R_0$) and the Bo number contribute the most to the uncertainty in the measured surface tension.

\subsection{Experimental procedure and analysis of the du No\"uy ring}

\begin{figure}[h]
    \centering
    \includegraphics[width=1\textwidth]{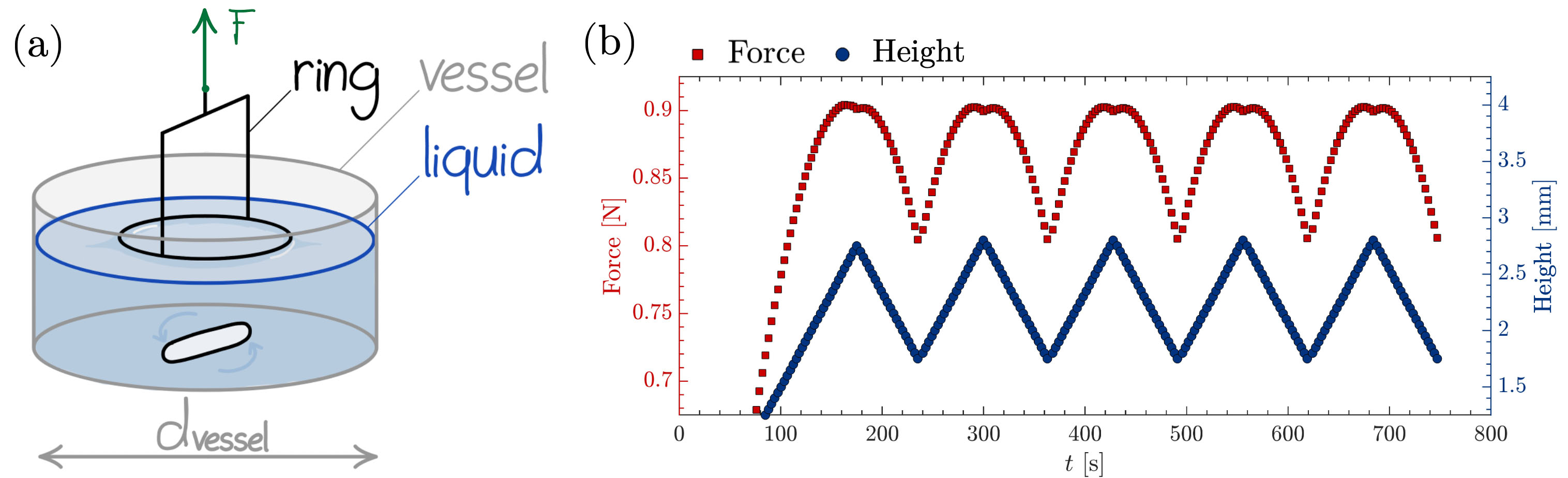}
    \caption{(a) Schematic of the du No\"uy ring surface tensiometer setup. (b) Typical measurement of the force over time as the height of the ring is changed.}
    \label{fig:setup_ring}
\end{figure}

The surface tension of water/1,2-hexanediol was measured using a force tensiometer (K100, KR\"USS) with a ring attachment (RI01, KR\"USS) with $d_\mathrm{ring} = 19.09$ mm and $r_\mathrm{wire} = 0.37$ mm. The platinum-iridium ring was thoroughly rinsed with ultra-pure water and any remaining residues were removed using a Bunsen burner, ensuring the ring was ideally wetting. At the start of the measurement series, 1,2-Hexanediol was placed in a glass vessel ($d_\mathrm{vessel} = 50$ mm) and the ring was gently lowered until submerged, see Fig. \ref{fig:setup_ring}(a). Then, the ring was pulled up until the maximum force was reached and lowered again. This was repeated five times per measurement, see Fig. \ref{fig:setup_ring}(a). The surface tension, including correction factor, was calculated by the accompanying software of the force tensiometer and probe (KR\"USS ADVANCE 1.11.0.15801).

To measure different concentrations, an automated dispensing system was used (Micro Dispenser, KR\"USS). The concentration of 1,2-hexanediol was gradually lowered by taking out some of the current mixture and replacing it with ultra-pure water. During the time the dispensers were active, a magnetic stirred ensured that the mixture was well mixed. After changing the composition, the next surface tension measurement was started after a 10 second pause, to let the flow subside. During all measurements, the temperature was actively controlled and kept constant at $23.00 \pm 0.07\, \mathrm{^\circ C}$.

\subsection{Calibration procedure of resistance temperature detector}

\begin{figure}[tbp]
    \centering
    \includegraphics[width=0.85\textwidth]{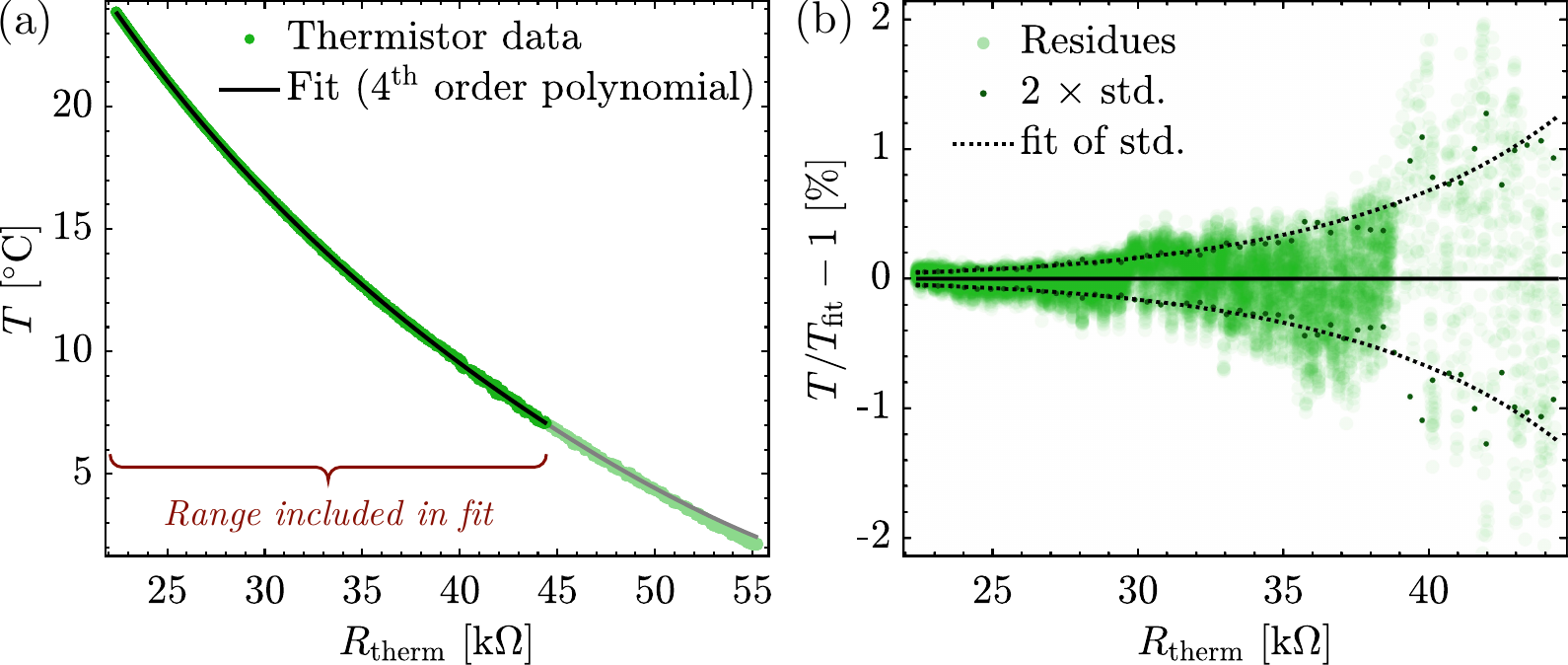}
    \caption{(a) The temperature measured by pt100 probe is shown versus the resistance measured by the thermistor. This curve is used to calibrate the thermistor with a fourth order polynomial function. Only the relevant temperatures ($> 7\, \mathrm{^\circ C}$) are included in the fit. (b) The difference between the measured temperature and the fitted temperature is shown as a function of the resistance. The standard deviation of the difference is calculate for short intervals as a function of the resistance and shown as dots. The standard deviation significantly increases with the resistance and is therefore fitted with a second order polynomial, which is shown as the discontinuous line.}
    \label{fig:cal_therm}
\end{figure}

A thermistor, also known as resistance temperature detectors, was used to measure the temperature inside the pendant drop while simultaneously measuring the surface tension with the pendant drop method. The thermistor (A96N4-GC11KA143L/37C, Amphenol Advanced Sensors) was sufficiently small (diameter = 0.3 mm) to fit through the needle and into the drop without disturbing the liquid-air interface. The thermistor was connected to a data acquisition unit (34970A, Keysight) that recorded the data to a PC \footnote{Source code thermistor (Python): \url{https://github.com/Dennis-van-Gils/project-thermistor-calibration}}. By appling a small current (10 $\tcmu$A) it was ensured that the heating due to thermal dissipation was minimal ($< 4 \ \mathrm{\tcmu W}$). 

The thermistor was calibrated by placing it in a temperature controlled liquid bath (PD15R-30-A12E, PolyScience) with an already calibrated thermometer (PT-104 pt100, Picotech) and measuring both the resistance and temperature while gradually decreasing the temperature of the liquid bath over the course of four hours. Fig. \ref{fig:cal_therm}(a) shows the temperature versus resistance and a fourth order polynomial fit of the data. Temperatures below $7\,\mathrm{^\circ C}$ were excluded as they were beyond the required temperature range and only introduced unnecessary noise. Fig. \ref{fig:cal_therm}(b) shows the difference between the measured temperature and the fitted temperature for the different resistances. The spread of the data increases with larger resistances. To quantify this, the standard deviation of the distribution was calculated on short intervals of resistance ($\approx 0.45 \ \mathrm{k\Omega}$). The binned standard deviation was fitted with a second order polynomial to accurately reflect increase in the uncertainty in the temperature with increasing resistance (or lower temperatures).

\subsection{Numerical simulations}

For the numerical simulations, we utilize our (open source) finite element framework \textsc{pyoomph} \cite{diddens2024jcp}, which is based on \textsc{oomph-lib} \cite{heil2006} and \textsc{GiNaC} \cite{bauer2002}. This framework was validated successfully against experiments in a plethora of publications, in particular on droplet evaporation \cite{diddens2021,raju2022,jalaal2022,rocha2024,diddens2024}.

By using a moving mesh, i.e. an arbitrary Lagrangian-Eulerian (ALE) approach, the moving liquid-gas interface is represented by a sharp interior boundary of the mesh. The mesh is based on the exact dimensions of the needle and the chamber. However, we assume axisymmetry in the simulations and therefore the cuboid shape of the chamber is converted to a cylinder with the same volume and height. In the liquid phase and optionally also in the gas phase, the Navier-Stokes equations are solved, where all fluid properties (mass density, viscosity, diffusivity, surface tension) vary with the local temperature and with the local composition based on fits of literature data \cite{takamura2012}. The pressure at the top opening of the needle is adjusted to obtain the desired volume of the pendant droplet. With two exceptions, no-slip boundary conditions are imposed at the walls of the needle and the container. In the vicinity of the pinned contact line, the no-slip condition is relaxed to a Navier-slip boundary condition to mediate the incompatibility of mass transfer and a pinned contact line. The particular value of the slip length ($\ell_\text{slip}=\SI{1}{\micro\meter}$) is chosen to be on the order of a typical element size near the contact line, but the specific choice of any reasonable value does not influence the overall results at all \cite{diddens2021}.

While the mass density is allowed to depend on the fluid composition and temperature, an explicit dependence on the pressure is not considered, i.e., the flow is incompressible, but not divergence-free. As a consequence of this in combination with mass transfer due to evaporation, the total fluid volume is not conserved, which requires to allow for leakage through the top boundary of the container, which we model by a strongly damped outflow, analogously to the Navier-slip condition, but in normal direction instead:
\begin{equation}
    2\mu \partial_z u_z-p = -\frac{\mu}{\ell_\text{leak}}u_z
\end{equation}
The influence of our choice of the leakage parameter $\ell_\text{leak}=\SI{1}{\micro\meter}$ is negligible, since it leads to typical leakage velocities of $10^{-7} $\si{\meter/\second} and corresponding pressure offsets of $10^{-6}$ \si{\pascal}.
At the liquid-gas interface, both Laplace pressure and thermal/solutal Marangoni effects are considered, which -- in combination with gravity and the kinematic boundary condition -- results in a dynamic shape evolution of the pendant droplet. If gas flow is taken into account, tangent component of the velocity is continuous at the interface, whereas Stefan flow and the vapor recoil pressure are considered in normal direction, however, without any noticeable influence on the results. 

Mass transfer is based on the Hertz-Knudsen-Schrage equation \cite{schrage1953}, which nearly instantaneously relaxes the vapor concentration to the vapor-liquid equilibrium, which we calculate based on the Antoine equation and by Raoult's law generalized by activity coefficients predicted by AIOMFAC \cite{zuend2008} in the case of liquid mixtures. The vapor concentration at the bottom of the container is either set to the saturated vapor concentration or to zero, depending of whether the corresponding experiment was carried out at high or low relative humidity. 
If thermal effects are considered, an advection-diffusion equation for the temperature field is solved inside both fluid domains and transient conduction is solved inside the needle, with continuous temperature at the interior boundaries and the experimentally measured chamber temperature at all exterior boundaries and as initial condition. At the liquid-gas interface, the latent heat of evaporation is considered.
 
As observables, we monitor the bulk-averaged temperature inside the drop and the surface-averaged surface tension. Due to the presence of flow and mass transfer and due to the varying mass density and surface tension, the shape of the droplet can deviate from the ideal solution of the Young-Laplace equation. We therefore also perform a fit of the surface tension by minimizing the residual of the Young-Laplace equation. Opposed to the experiments, where the measured interface shape has been fitted iteratively by the Young-Laplace equation, it is possible to minimize the functional
\begin{equation}
    F = \int_S \left(\gamma_\text{fit}\kappa +\rho g z-\Delta p\right)^2\: dS
\end{equation}
with curvature $\kappa$ with respect to $\gamma_\text{fit}$ and $\Delta p$ along with all other equations during the simulation. The integral is carried out over the numerically obtained dynamic interface shape. This procedure yields $\gamma_\text{fit}$ and its standard deviation automatically at each time step.
 
More details on the simulation method (e.g. weak formulations) can be found in the supplementary material of Ref. \cite{jalaal2022} and the documentation of \textsc{pyoomph} \footnote{Documentation of \textsc{pyoomph}: \url{https://pyoomph.github.io}}. The source code of the simulations is made available on request.


\section{Effect of evaporation on the pendant drop method} \label{sec:evaporation}

\begin{figure}[bp]
    \centering
    \includegraphics[width=0.8\textwidth]{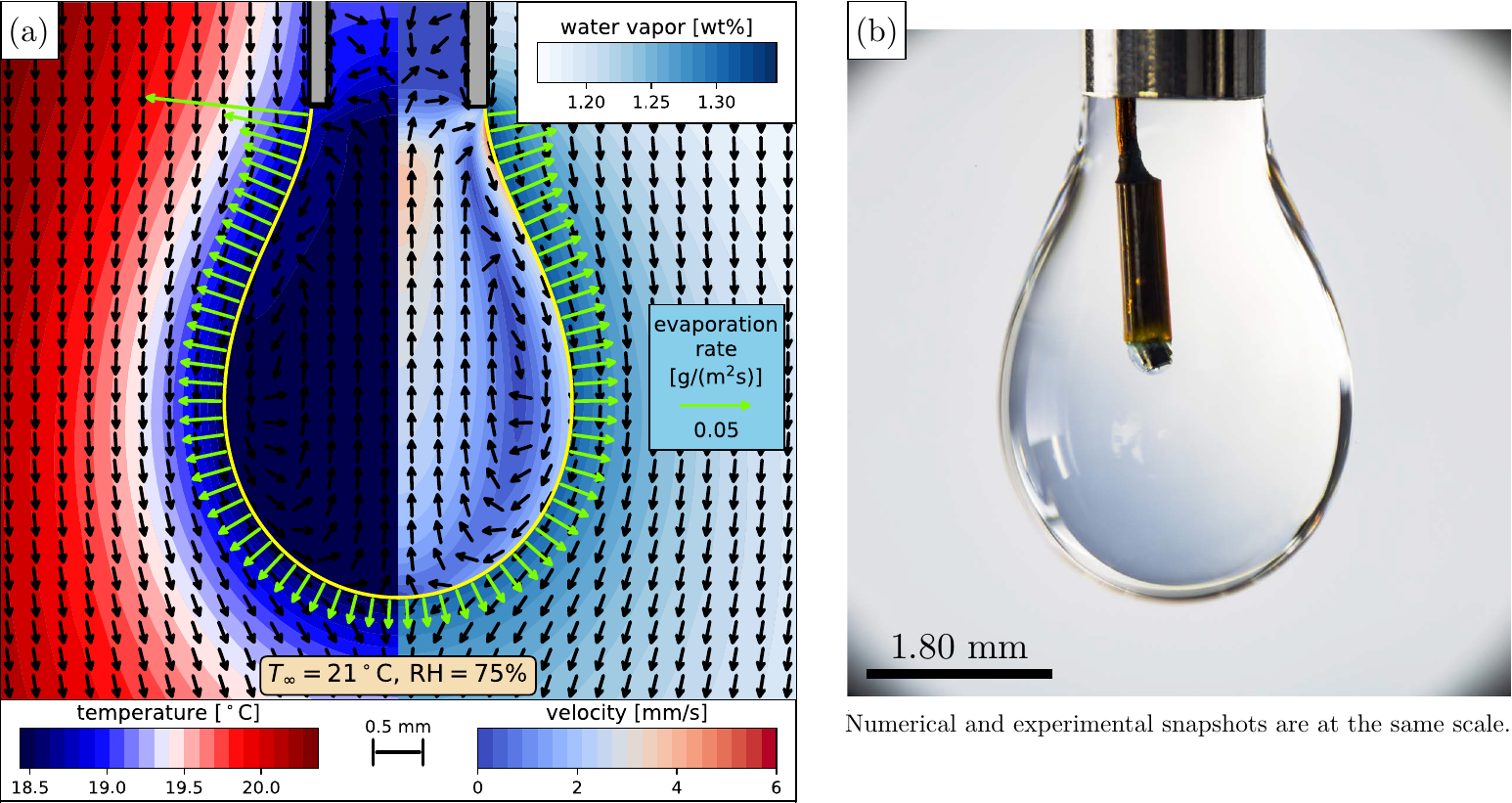}
    \caption{Snapshot of an evaporating pendant drop. (a) Numerical simulation of a water drop at 75$\%$ relative humidity. The left half shows the temperature in and around the drop. The right half shows the velocity in the liquid and the air. The green arrow perpendicular to the drop surface indicate the local evaporative flux. (b) Experiment of a water drop at 100$\%$ relative humidity. The thermistor inside the drop becomes clearly visible when using a diffuser instead of an aperture for background lighting. The scales of (a) and (b) are identical.} 
    \label{fig:num_and_exp}
\end{figure}

\begin{figure}[bp]
    \centering
    \includegraphics[width=1\textwidth]{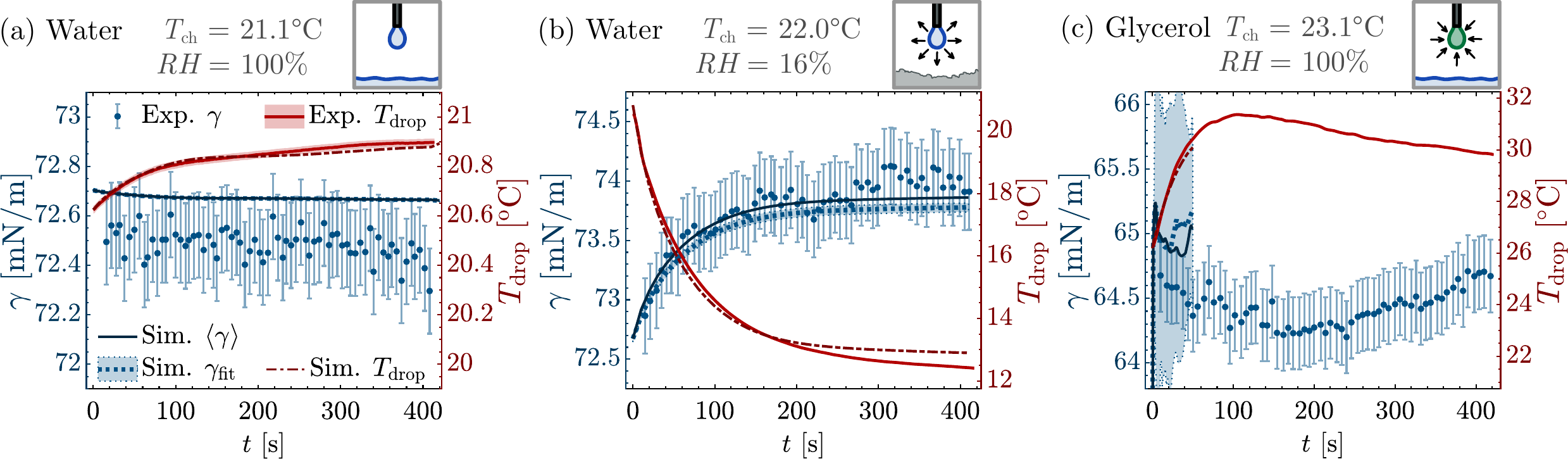}
    \caption{The surface tension and temperature of a pedant drop subject to different relative humidities: (a) A water drop at high relative humidity, (b) a water drop at low relative humidity, (c) a glycerol drop at high relative humidity. Both the experimentally measured surface tension with the pendant drop method and the temperature measured with the thermistor are shown as a function of time. Similarly, the average temperature and the average surface tension, $\langle\gamma\rangle$, from the numerical simulations are shown as a function of time. Additionally, to mimic the pendant drop method, the apparent surface tension, $\gamma_\mathrm{fit}$, when fitting the numerically calculated drop shape with the Young-Laplace (equation \ref{eq:fit_YL_eq}), is shown. The uncertainties (standard deviation) are shown by the shaded regions. The error bars show the uncertainty in the experimental surface tension measurement. The ambient temperature and relative humidity for each measurement are shown in Fig. \ref{fig:sensors_all}.}
    \label{fig:gamma_and_temp}
 
    \vspace{12 pt}

    \centering
    \includegraphics[width=1\textwidth]{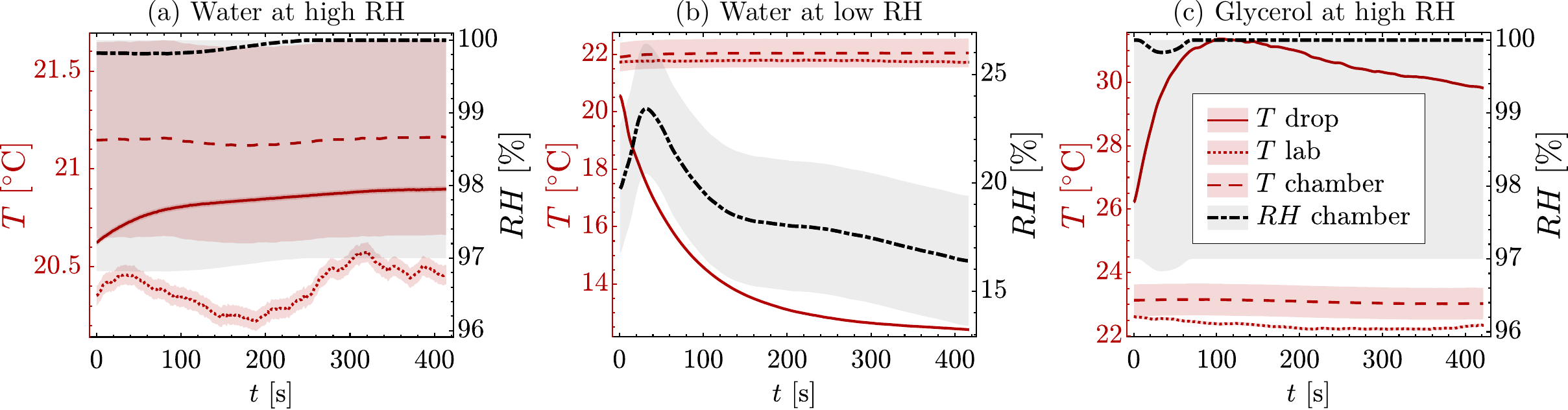}
    \caption{The temperature and relative humidity shown in the panels (a), (b), and (c) correspond to the pendant drop measurements shown in Fig. \ref{fig:gamma_and_temp}. The measured temperature and relative humidity in the chamber and the temperature in the lab (outside the chamber) are shown as a function of time. For reference, the temperature measured in the drop is also shown. The uncertainty in each measurement is shown by a shaded region. The legend given in (c) holds for all three panels.}
    \label{fig:sensors_all}
\end{figure}

First we investigate the effect of evaporative cooling by measuring the temperature in the drop using a thermistor and simultaneously using the pendant drop method to calculate the surface tension of the drop, see Fig. \ref{fig:num_and_exp}(b). We can control the evaporation by performing the pendant drop method in different relative humidities. Fig. \ref{fig:gamma_and_temp}(a) shows the temperature and measured surface tension over time of a pendant water drop at high relative humidity ($RH>99.5\%$). The temperature slightly increases ($\Delta T_\mathrm{drop}\approx 0.3 \mathrm{^\circ C}$) due to a difference in the temperature of the newly produced drop and the temperature inside the chamber. The measured surface tension remains constant. 

In contrast, for a pendant water drop under the same conditions but at a low relative humidity ($RH \rightarrow 16\%$), very different behavior is observed, as is shown in Fig. \ref{fig:gamma_and_temp}(b). Firstly, the temperature in the drop reduces drastically due to evaporative cooling. After only 10 s the temperature drop is lowered by $2.6 \mathrm{^\circ C}$, then, after few minutes, the temperature stabilises with a total temperature difference $\Delta T_\mathrm{drop} \approx - 9.5 \mathrm{^\circ C}$. Secondly, the measured surface tension increases sharply and then also stabilises having increased a total of $\Delta\gamma \approx 1.5 \ \mathrm{mN/m}$.

During the pendant drop measurements shown in Fig. \ref{fig:gamma_and_temp}, the relative humidity in the chamber, the temperature in the chamber, and the temperature in the lab (outside the chamber) were measured and are shown in Fig. \ref{fig:gamma_and_temp}. For the water drop at high relative humidity, we can verify that the relative humidity is nearly 100$\%$ and that the temperature difference inside the drop are in between the lab temperature and the chamber temperature. For the water drop at low relative humidity, the relative humidity is not constant but varies over time. Initially, it increases to a maximum of $23.4\%$ and then reduces again. Since the humidity sensor is positioned near the top of the chamber relatively close to the drop, we measure the increased relative humidity around the evaporating drop, while the hygroscopic beads are at the bottom of the chamber. As the drop cools down, the evaporation considerably reduces since the vapor pressure of water depends strongly on temperature \cite{huang2018}. After a few minutes, the relative humidity is still slowly decreasing, approaching $RH=16\%$.

For a hygroscopic liquids of mixtures at high relative humidity, the opposite of evaporative cooling, heating due to condensation of water vapor, can be relevant. This is exemplified by a pendant glycerol drop at high relative humidity, which is shown in Fig. \ref{fig:gamma_and_temp}(c). The temperature of the drop initially increases and reaches a maximum of $31.4\mathrm{^\circ C}$ at $t=100$ s. After that the drop slowly cools down. The measured surface tension initially decreases and reaches a minimum at $t=200$ s before steadily increasing. Whereas the temperature and surface tension are corresponding to each other in time for water at low relative humidity, for glycerol at high relative humidity, the maxima in temperature occurs at a different time than the minimum in surface tension. This is because the surface tension is affected by two competing mechanisms. 1: Due to the temperature increase of the drop, the surface tension will decrease. 2: Due to the increasing water concentration of the drop, the surface tension will increase.

To further elucidate the effect of evaporation on the measured surface tension beyond evaporative cooling we use numerical simulations, where we take all relevant effects into account such as: non-uniform evaporative flux, thermal effect, flow in the drop and the gas phase, and deformation of the pendant drop shape. Fig. \ref{fig:num_and_exp} shows a numerical snapshot of an evaporating pendant water drop next a experimental snapshot. By simulating the exact experimental conditions, including the finite chamber size, the needle dimensions, the reservoir at the bottom of the chamber, and the initial temperature of the drop we find good agreement between the experiment and the numerics for pendant water drops, see Fig. \ref{fig:gamma_and_temp}. For glycerol at high relative humidity, no good agreement was found, owing to several reasons. Most importantly, the flow is no longer axisymmetric since the solutal Marangoni flow is unstable \cite{gelderblom2022}. Additionally, some parameters are unknown, such as the latent heat of water for a binary mixture of water and glycerol.

\begin{figure}[bp]
    \centering
    \includegraphics[width=0.8\textwidth]{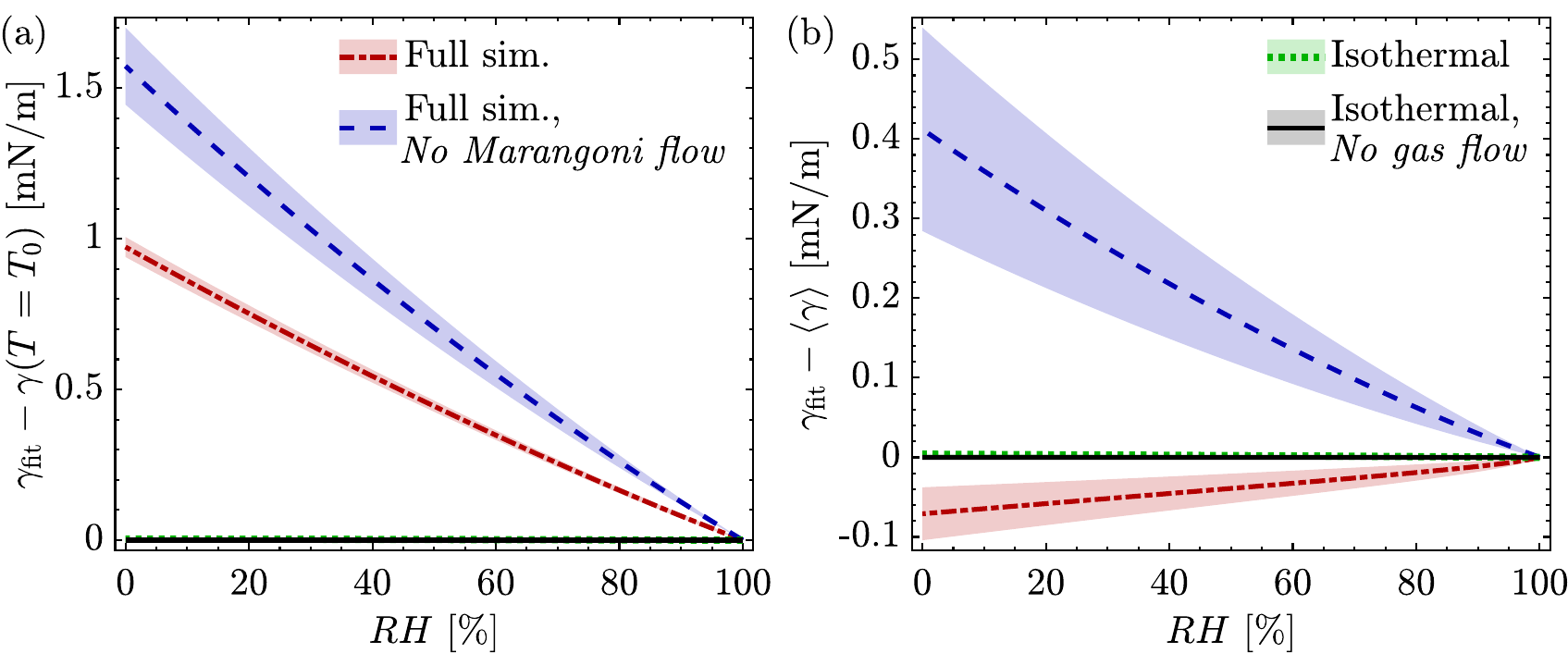}
    \caption{(a) Difference between the apparent surface tension, $\gamma_\mathrm{fit}$, (YL-fit of drop shape) and the initial surface tension, $\gamma(T = T_0)$, of the drop versus the relative humidity. (b) Difference between the apparent surface tension and the average surface tension, $\langle\gamma\rangle$, versus the relative humidity. Numerical simulations with different effects are shown: Isothermal (no evaporative cooling), with and without gas flow, full sim. (with evaporative cooling), with and without thermal Marangoni flow. For all simulations the steady state is shown. The shaded region corresponds to the uncertainty in the surface tension due to the YL-fit of the drop shape.}
    \label{fig:gamma_num_diff}
\end{figure}

In the numerics we find that the thermal gradients in the bulk of the drop are minimal (see Fig. \ref{fig:num_and_exp}). Hence, we expect that the experimentally measured surface tension is equivalent to the average temperature of the drop. However, the local surface tension is not constant on the surface of the drop. In particular close to the needle, where thermal gradients are more pronounced due to the heat supplied through the metal needle walls, resulting in a 0.5$\mathrm{^\circ C}$ temperature difference in the upper 15$\%$ of the needle. Moreover, the drop shape is not in equilibrium due to flow in the drop and in the air, which are caused by density and surface tension gradient driven flows. To quantify the relevance of these effects, we numerically evaluate the average surface tension, $\langle\gamma\rangle$, over the drop surface and compare that with the apparent surface tension, $\gamma_\mathrm{fit}$, that is obtained when fitting the drop shape with Young-Laplace equation \ref{eq:fit_YL_eq}. Both the average and fitted surface tension are shown in Fig. \ref{fig:gamma_and_temp} and both agree well with the experimental data, indicating that surface tension based on the YL eq. fit to the drop shape, $\gamma_\mathrm{fit}$, approximates the average surface tension, $\langle\gamma\rangle$, well.

Next, we systematically explore the difference between the apparent surface tension (YL-fit of drop shape) and the initial surface tension and the average surface tension for different relative humidities when the drop has reached a steady state, which is shown in Fig. \ref{fig:gamma_num_diff}. We investigate the relevance of different mechanisms by disabling different effects in the numerical simulations. Starting with an isothermal simulation (latent heat = 0) such that the evaporative cooling no longer plays a role. We see that the deviation from both the initial surface tension and the average surface tension is negligible (no gas flow: $< 5 \cdot10^{-5}$ mN/m, with gas flow: $<4 \cdot10^{-3}$ mN/m). 

However, when we enable evaporative cooling (Full sim. in Fig. \ref{fig:gamma_num_diff}), the apparent surface tension deviates by as much as $+1$ mN/m from the initial surface tension and by almost $-0.1$ mN/m from the average surface tension. Interestingly, when we disable thermal Marangoni flow, the deviation for both increases by approximately 0.5 mN/m. This is because the thermal Marangoni flow contributes significantly to convection of heat throughout the drop, homogenizing the temperature at the drop surface from 5.73$\mathrm{^\circ C}$ without thermal Marangoni flow to 1.66$\mathrm{^\circ C}$ with thermal Marangoni flow. Consequently, without thermal Marangoni flow, the surface tension gradients are significantly stronger, resulting in a worse fit of the YL equation to the drop shape.

In conclusion, we observe both in experiments and numerics that evaporative cooling strongly impacts the outcome of a pendant drop measurements. Additionally, evaporation induces surface tension gradients and flow in the drop, altering the drop shape from the shape given by the Young-Laplace equation \ref{eq:fit_YL_eq}, which is valid only when the drop is in equilibrium (no flow, constant surface tension). As a result of the non-ideal drop shape, the apparent surface tension that is measured is typically lower than the average surface tension. However, this effect is much smaller compared to the increase in surface tension due to evaporative cooling.


\section{Surface tension of binary aqueous mixtures of n-diols and glycerol} \label{sec:surface_tension}

\begin{figure}[bp]
    \centering
    \includegraphics[width=0.85\textwidth]{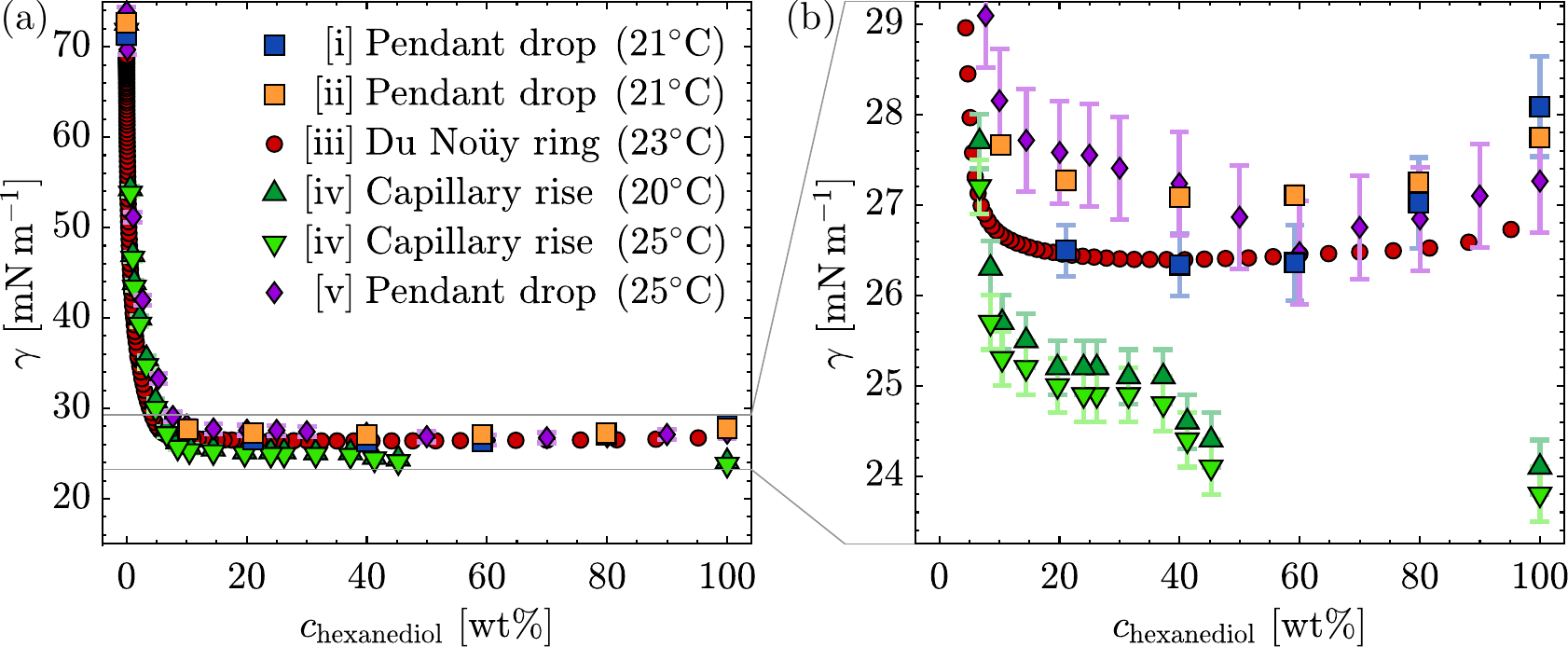}
    \caption{Surface tension of water/1,2-hexanediol as a function of composition measured using different techniques as indicated by the legend. Note that the measurements were performed at different temperatures. (a) Shows the whole range. (b) is zoomed in. The following data is shown: [i] This work (\textit{initial estimate}); [ii] This work (\textit{refined measurement}), the error bars are similar in size as the markers; [iii] This work, no error bars are shown; [iv] \citet{romero2007fpe}, at two different temperatures; [v] \citet{hack2021}.}
    \label{fig:gamma_hd}
\end{figure}

Now we turn to the case of water/1,2-hexanediol, which is of particular interest due to the remarkable segregation dynamics that is observed when a sessile drop of water/1,2-hexanediol evaporates. In the complementary paper to this work \cite{diddens2024}, numerical simulations show that the segregation dynamics can be understood if the surface tension of water/1,2-hexanediol in non-monotonic. Having established that evaporation can significantly change the outcome of a pendant drop measurement, from now on we will perform all pendant drop measurements at equilibrium vapor pressure to minimise any effect of evaporation, unless otherwise specified.

All the data on the surface tension of water/1,2-hexanediol are shown together in the centre figure of the paper, Fig. \ref{fig:gamma_hd}, these are:
\setlist{nolistsep} 
\begin{enumerate}[noitemsep]
    \item[\text{[i]}] The surface tension  measured using the pendant drop method (this work). However, this was an initial estimate and no humidity control was used. Additionally, during the measurement some vibrations (either due to airflow or the setup itself) were observed and no correction for the camera angle was applied. This is (partially) reflected by the larger uncertainties. 
    \item[\text{[ii]}] The surface tension  measured using the pendant drop method  (This work). The uncertainties are roughly the same size as the markers. 
    \item[\text{[iii]}] The surface tension  measured using the du No\"uy ring method (This work). 
    \item[\text{[iv]}] The surface tensio  measured using the capillary rise method by \citet{romero2007fpe} using equation \ref{eq:gamma_cap}. It is shown for two different temperatures, $20\mathrm{^\circ C}$ and $25\mathrm{^\circ C}$. Since, in that work, the focus was on dilute solutions, no data are provided for concentrations larger than $c_\mathrm{hexanediol} = 45\ \mathrm{wt\%}$.
    \item[\text{[v]}] The surface tension measured using the pendant drop method by \citet{hack2021}. In that work no humidity control was used.
\end{enumerate}
    
Starting with the overall surface tension of water/1,2-hexanediol, see Fig. \ref{fig:gamma_hd}(a), we find that this shape is characteristic of the surface tension of a surfactant. Initially, the decrease in surface tension with increasing concentration is incredibly steep. But from a certain concentration onward the surface tension is nearly constant, which is known as the critical micelle concentration (CMC) for surfactants since from this point onward micelles start to form in the bulk of the liquid when more surfactant is added. In this case the equivalent CMC for water/1,2-hexanediol is roughly 6 wt$\%$.

Although the surface tension appears constant for larger concentrations, when we inspect the surface tension for water/1,2-hexanediol beyond $c_\mathrm{hexanediol} = 6\ \mathrm{wt\%}$ as shown in Fig. \ref{fig:gamma_hd}(b), we find that the surface tension is indeed non-monotonic according to four out of five data sets. Despite the relatively large spread of the data, every pendant drop measurement, and the du No\"uy ring measurement show an increase in the surface tension and a minimum at approximately $c_\mathrm{hexanediol} \approx 40\ \mathrm{wt\%}$. Only the capillary rise measurement does not show this increase, however, as mentioned before, no data points exist between $45\ \mathrm{wt\%}$ and $100\ \mathrm{wt\%}$. Therefore, no definitive conclusion can be made based on this capillary rise measurement.

We emphasise that the non-monotonic dependence on composition of the surface tension of water/1,2-hexanediol is very unusual, particularly considering that 1,2-hexanediol is not a complex molecule from a chemical perspective and is fully miscible in all ratios in water. To further investigate, we measure the surface tension of aqueous binary mixtures of 1,2-butanediol and 1,2-pentanediol. Additionally, we also include 1,5-pentanediol which also has an unusual surface tension \cite{glinski2000}. As an extra verification of our pendant drop method we also include water/glycerol, which has a well established surface tension in the literature \cite{takamura2012}. 

\begin{figure}[tp]
    \centering
    \includegraphics[width=0.85\textwidth]{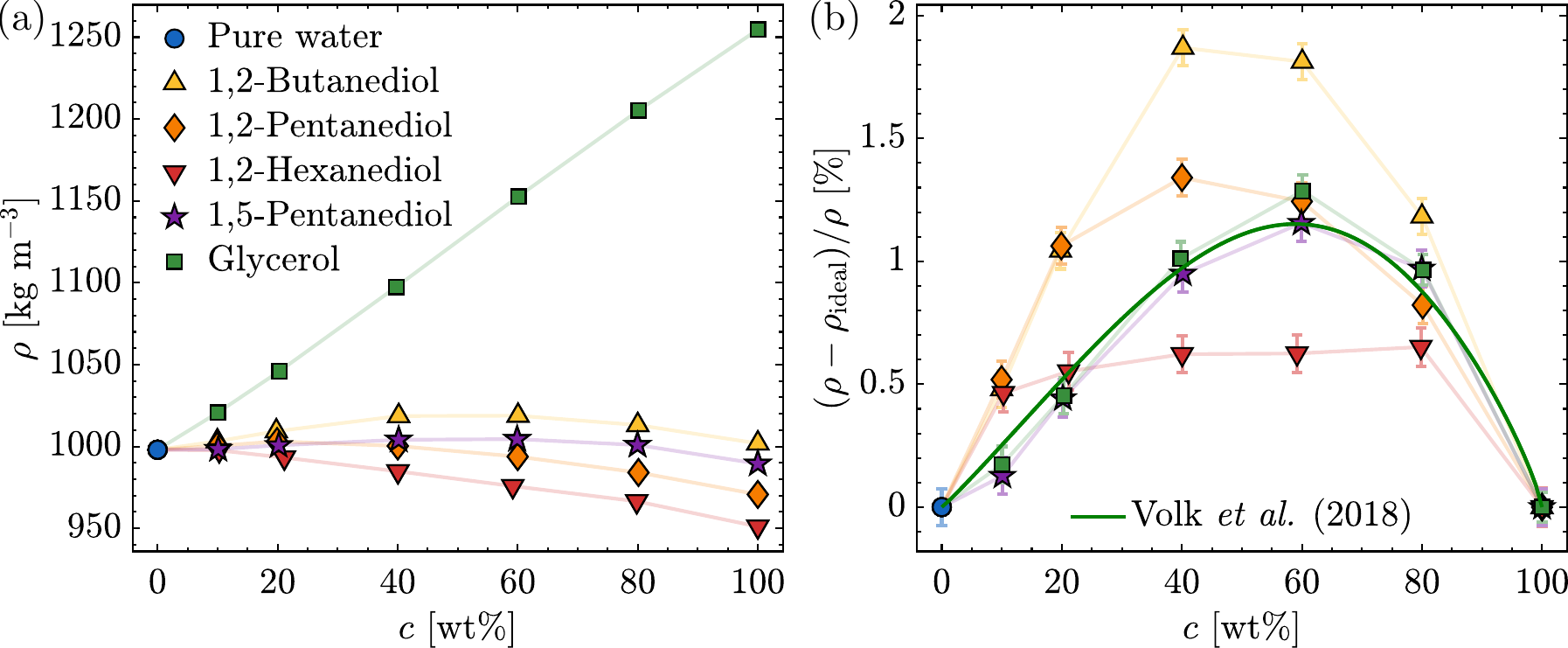}
    \caption{(a) The density of binary aqueous mixtures of n-diols and glycerol. (b) The normalised deviation of the measured density from the density for an ideal mixture given by equation \ref{eq:rho_ideal}, which is equivalent to the volume contraction coefficient. For reference \cite{volk2018} is shown.}
    \label{fig:density_all}

\vspace{12 pt}

    \centering
    \includegraphics[width=0.8\textwidth]{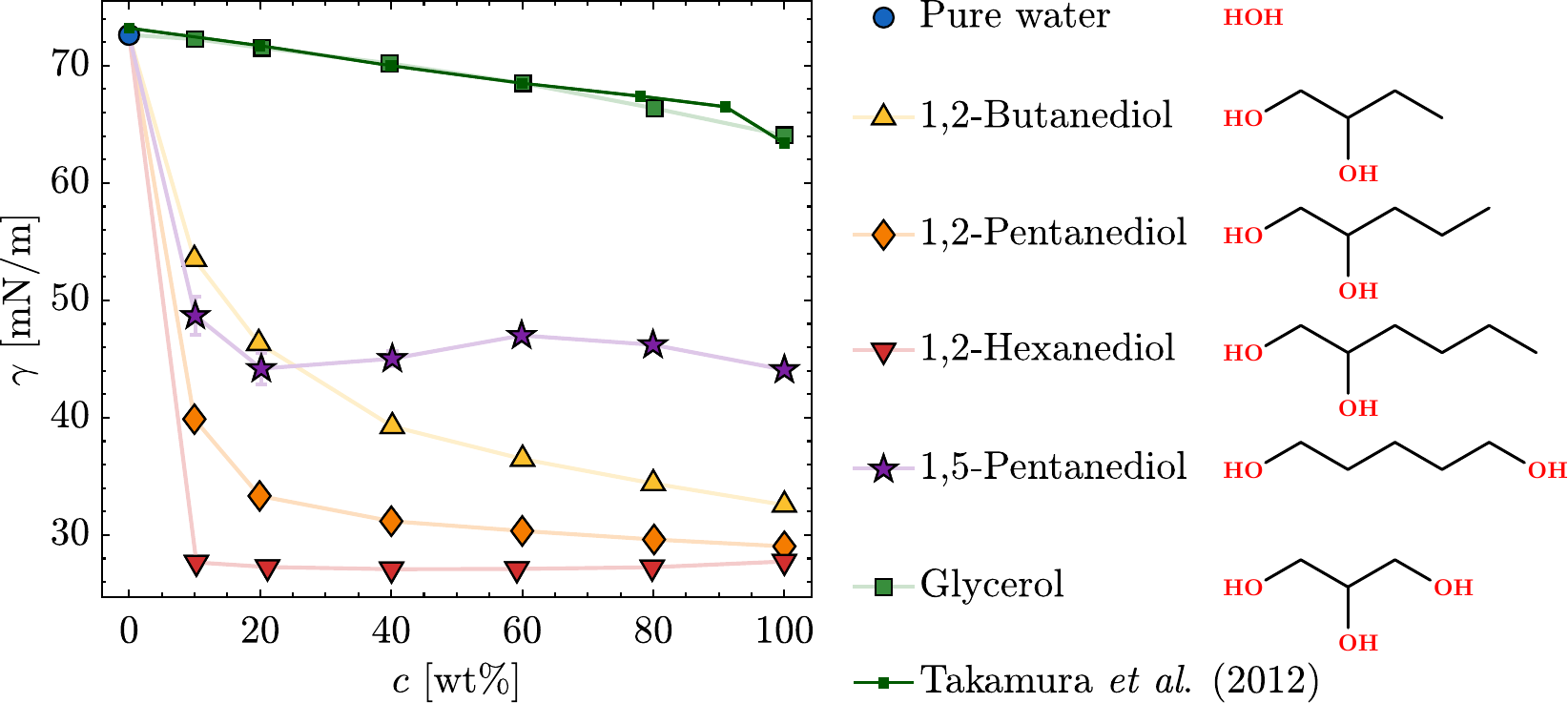}
    \caption{Surface tension of binary aqueous mixtures of n-diols and glycerol. For reference, the surface tension of water/glycerol measured by \cite{takamura2012} is shown.}
    \label{fig:gamma_all}
\end{figure}

First, the density was measured using an oscillating U-tube (DMA 35, Anton Paar) of all the aqueous mixtures. The results are shown in Fig. \ref{fig:density_all}(a). While the densities of water/glycerol and water/1,2-hexanediol are monotonic increasing, the densities of the other mixtures are non-monotonic and have a maximum density that is larger than either of the pure components. Therefore, we compare the measured densities to the density of an ideal mixture, $\rho_\mathrm{ideal}$, which is given by:
\begin{equation}
    \rho_\mathrm{ideal} = \rho_a\phi_a + \rho_b\phi_b = \frac{\rho_a \rho_b}{\rho_a c_b +  \rho_b c_a}
    \label{eq:rho_ideal}
\end{equation}
Here, $\rho_a$ and $\rho_b$ are the densities of the pure components $a$ and $b$ respectively, $\phi_a$ and $\phi_b$ are the volume fractions, and $c_a$ and $c_b$ are the mass fractions. The normalized difference between the measured and the ideal density is shown in Fig. \ref{fig:density_all}(b). This is equivalent to the volume contraction coefficient, i.e., how much the volume of a mixture reduces compare to the volumes of the separate unmixed components. All mixtures show a positive volume contraction coefficient. For comparison a fit by \citet{volk2018} based on literature values for water/glycerol is shown that agrees well with the experimental data.

With the density data and the fitted Bo number from the pendant drop method we can compute the surface tensions for all mixtures. These are shown in Fig. \ref{fig:gamma_all}.  Starting with the surface tension of water/1,2-butanediol, the surface tension decrease for low concentrations is strong and becomes more gradual for large concentration. Moving on to water/1,2-pentanediol, the slope is considerable steeper for low concentrations and flattens out more for large concentrations, resembling the surface tension for water/1,2-hexanediol more closely. As discussed before, water/1,2-hexanediol has an even steeper slope for low concentrations, similar to a surfactant and for large concentrations the surface tension appears constant at first glance, but in fact increases slightly. As validation of the surface tension measurements, literature values are shown for water/glycerol from \cite{takamura2012}, which agree well with our results.

The surface tension for water/1,5-pentanediol is even more exotic than for water/1,2-hexanediol. For low concentrations the surface tension steeply decreases, similar to water/1,2-pentanediol. Then, the surface tension has a minimum around $c_\mathrm{1,2-pentanediol} \approx 30 \ \mathrm{wt\%}$ and increases again. Then, the surface tension has a maximum as well around $c_\mathrm{1,2-pentanediol} \approx 60 \ \mathrm{wt\%}$, before decreasing again for the remaining concentrations. This S-shaped surface tension depending on the composition has been observed before by \cite{glinski2000}, who excluded impurities as the cause of this dependence. Here, we observe that mixtures of water/1,5-pentanediol, for concentrations between 10 wt$\%$ and 40$\%$, are visibly cloudy, suggesting that the mixture is not perfectly miscible. For large concentrations larger than 40$\%$ the solutions are fully transparent, and therefore appear to be fully miscible again. Additionally, for concentrations between 10 wt$\%$ and 40$\%$, the surface tension decreased significantly over time (in the order of 5 mN/m over tens of seconds), while controlling the humidity to minimize the effect of evaporation. 

We speculate that the molecular structures formed by water/1,5-pentanediol are very large compared to typical binary mixtures. When suddenly an interface is created, this changes the structure and possibly even the local distribution of water/1,5-pentanediol, which could take a long time if the structures are sufficiently large. However, beyond this speculation, no conclusions can be made. Further investigation of the time-dependence of the surface tension and the molecular structure using molecular dynamics simulation might provide insight in the underlying mechanism of the surface tension of water/1,5-pentanediol. Similarly, we can speculate that the molecular structure to plays a critical role in the non-monotonic surface tension dependence on composition of water/1,2-hexanediol. However, we defer the explanation to future investigations.


\section{Conclusion and outlook} \label{sec:conclusion}

In summary, we have experimentally and numerically studied the effect of evaporation on the measured surface tension by the pendant drop method. Using a thermistor and an evaporation controlled setup, we experimentally measured the temperature and surface tension simultaneously. We found that evaporative cooling drastically lowers the temperature in the drop due to evaporative cooling, by as much as 9.5$\mathrm{^\circ C}$, resulting in a significantly higher measured surface tension. Using numerical simulations, in addition to finding good agreement with the experimental results, we investigate to what extend the apparent surface tension of the drop is affected by the shape deviation from the equilibrium shape (with constant surface tension and no flow) to the actual shape of the drop at different relative humidities. We found that there is some deviation between the apparent surface tension (based on the fit of the drop shape) and the average surface tension over the drop surface, but that the effect of evaporative cooling is much more important. Therefore, controlling evaporation is essential for accurate surface tension measurements using the pendant drop method. Additionally, we have provided a detailed description of various other experimental aspects concerning the accuracy of the pendant drop method that can be used for future reference.

Next, we applied the pendant drop method to measure the surface tension of water/1,2-hexanediol and compared it to various other methods and literature values. A non-monotonic dependence with increasing concentration of 1,2-hexanediol was found with a minimum surface tension around $c_\mathrm{hexanediol} = 40 \ \mathrm{wt\%}$. Although the surface tension increase is very small $(\approx 0.5 \ \mathrm{mN/m})$, this is sufficient to drastically alter the segregation dynamics in an evaporating sessile water/1,2-hexanediol drop, which is investigated in detail in the counterpart of this work, \cite{diddens2024}. In an attempt to obtain further insight into the origin of the non-monotonic dependence, the surface tension for various other n-diols are also measured. Although we do see a tendency to more surfactant-like surface tensions from 1,2-butanediol to 1,2-hexanediol, the underlying mechanism remains elusive and subject to future investigation, including molecular dynamics simulations.

The results presented in this work are pertinent to any applications or processes in which accurate and precise measurements of the surface tension are imperative, particularly for evaporation driven processes, which often involve surface tension gradients.


\section{Acknowledgments}

This work was supported by an Industrial Partnership Programme, High Tech Systems and Materials (HTSM), of the Netherlands Organisation for Scientific Research (NWO); a funding for public-private partnerships (PPS) of the Netherlands Enterprise Agency (RVO) and the Ministry of Economic Affairs (EZ); Canon Production Printing Netherlands B.V.; and the University of Twente. The authors thank Hank-Jan Koier for his help with the surface tension measurement using the du No\"uy ring method.


\appendix

\section{Calibration}

\begin{figure}[b!]
    \centering
    \includegraphics[width=0.65\textwidth]{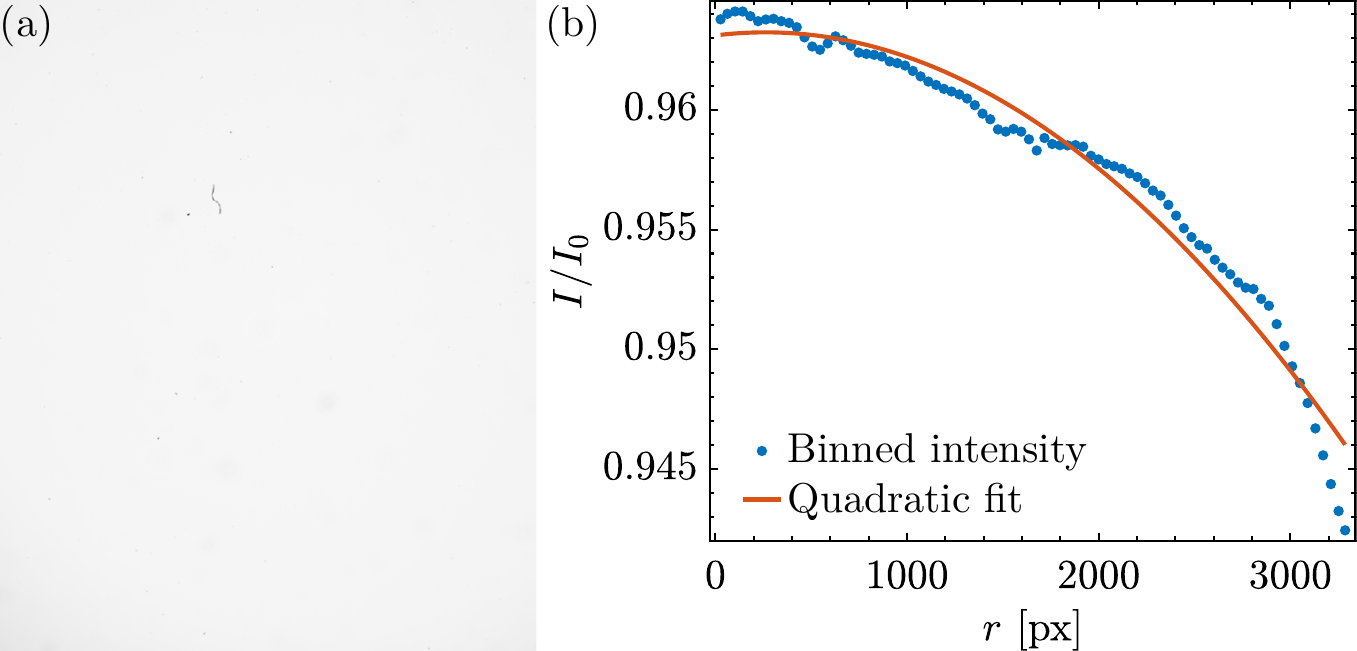}
    \caption{Calibration of the background illumination and lens vignetting. (a) Blanc image without the drop. (b) Intensity, normalised by the maximum intensity, as a function of distance from the optical axis. The difference in intensity between the centre and the edge of the image is approximately 2$\%$. The intensity drop is well approximated with a quadratic fit.}
    \label{fig:light}
\end{figure}

A proper calibration is essential for the pendant drop method since the calculated surface tension depends quadratically on the drop size $R_0$ cf. eq. \ref{eq:gamma_bo}. First, we calibrate the background illumination and lens vignetting to correct for any inhomogeneities across the image, which is important since the drop edge is determined using the threshold method. Fig. \ref{fig:light} shows an image without any object except the background and a quadratic fit of the intensities as a function of distance from the optical axis. In this case the illumination is very homogeneous and lens vignetting minimal, with roughly a 2$\%$ difference between the image centre and the corners.

\begin{figure}[h!]
    \centering
    \includegraphics[width=1\textwidth]{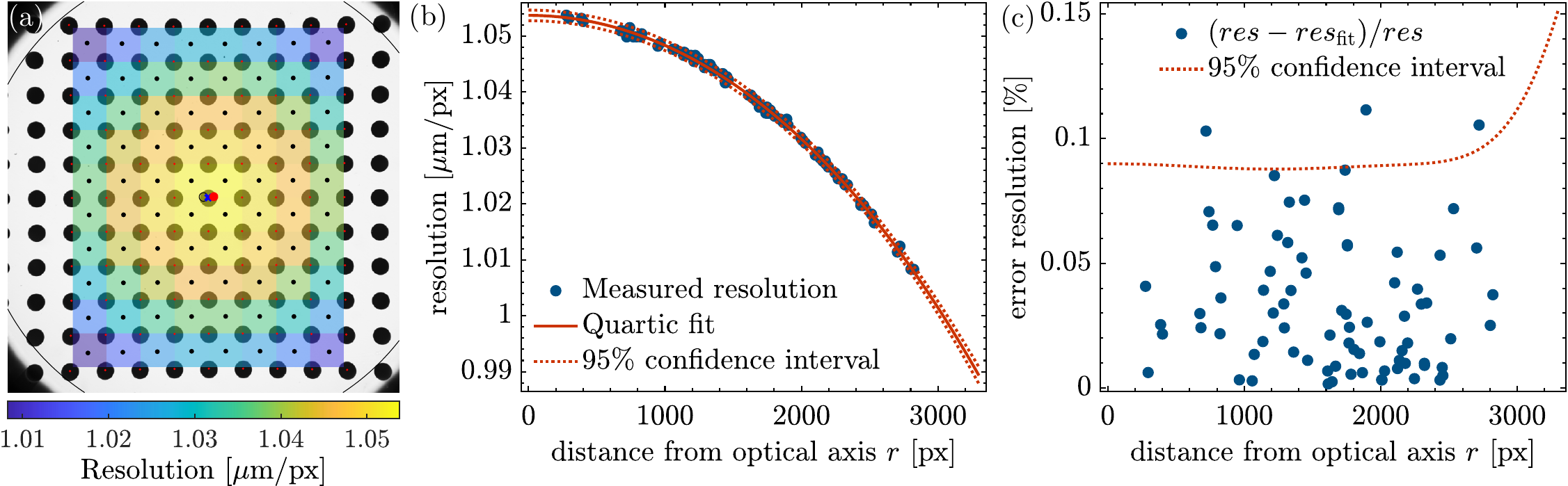}
    \caption{Calibration of the resolution and distortion of the camera and lens. (a) The image with the resolution target is shown with the local resolution superimposed. The the small red dots mark the centres of each patch on the resolution target. The black dots mark the positions where the resolution is evaluated. The symbols in the centre mark the image centre (open circle), the resolution target centre (blue cross), and the optical axis (large red circle). (b) The local resolution as a function of distance from the optical axis. A quartic fit of the data is shown with a 95$\%$ confidence interval. (c) The absolute value of the deviation between the fit and the data, normalised by the local resolution.}
    \label{fig:resolution}
\end{figure}

\begin{figure}[h!]
    \centering
    \includegraphics[width=1\textwidth]{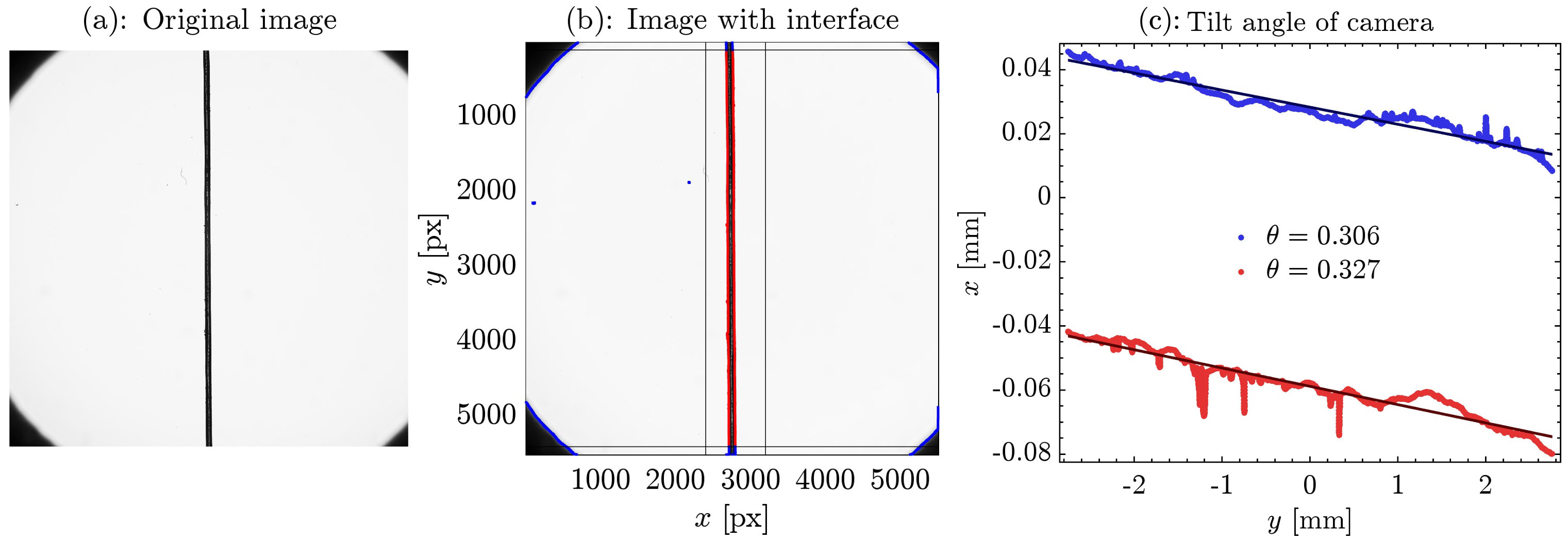}
    \caption{Camera tilt angle calibration. (a) Original captured image of pendulum. (b) Image corrected for background illumination and vignetting. The dots mark the edges in the image. Only the edges within the guide lines are considered when determining the camera angle and are coloured red. (c) The edges of either side of the pendulum is fitted with a line and the slope gives the camera tilt angle.}
    \label{fig:angle}
\end{figure}

Next, we calibrate the resolution of the camera and lens by imaging a grid distortion target (58-774, Edmund Optics), with known size. By convoluting the image using the center spot as a kernel we could accurately determine the relative positions of the spots in the image. The local resolution is determined by calculating the distance between the spots centers and combining this with the known distance between the spots (0.5 mm). Fig. \ref{fig:resolution}(a) shows the captured image of the resolution target with an overlay of the local resolution. The resolution was not constant, but is larger in the center and smaller in the corners of the image, resulting in pincushion distortion.

Fig. \ref{fig:resolution}(b) shows the local resolution as a function of distance from the optical axis. The data was fitted to a quartic fit (i.e. $y = a_0 + a_1 x^2 + a_2 x^4$) to account for the distortion. The fit seems to describe the data well. However, for this particular camera-lens combination, the center of the optical axis did not coincide with the center of the camera sensor. So in addition to the quartic fit, the position of the optical axis was also fitted. Fig. \ref{fig:resolution}(c) shows the deviation from the fitted resolution and the measured resolution.

Finally, we calibrate the tilt angle of the camera by imaging a pendulum and calculating the slope, as shown in Fig. \ref{fig:angle}. Each calibration was repeated for every measurement series to account for changes in lighting conditions, camera position, and zoom factor for the different drop sizes.

\FloatBarrier

\section{Fitting the pendant drop}

\begin{figure}[t!]
    \centering
    \includegraphics[width=0.75\textwidth]{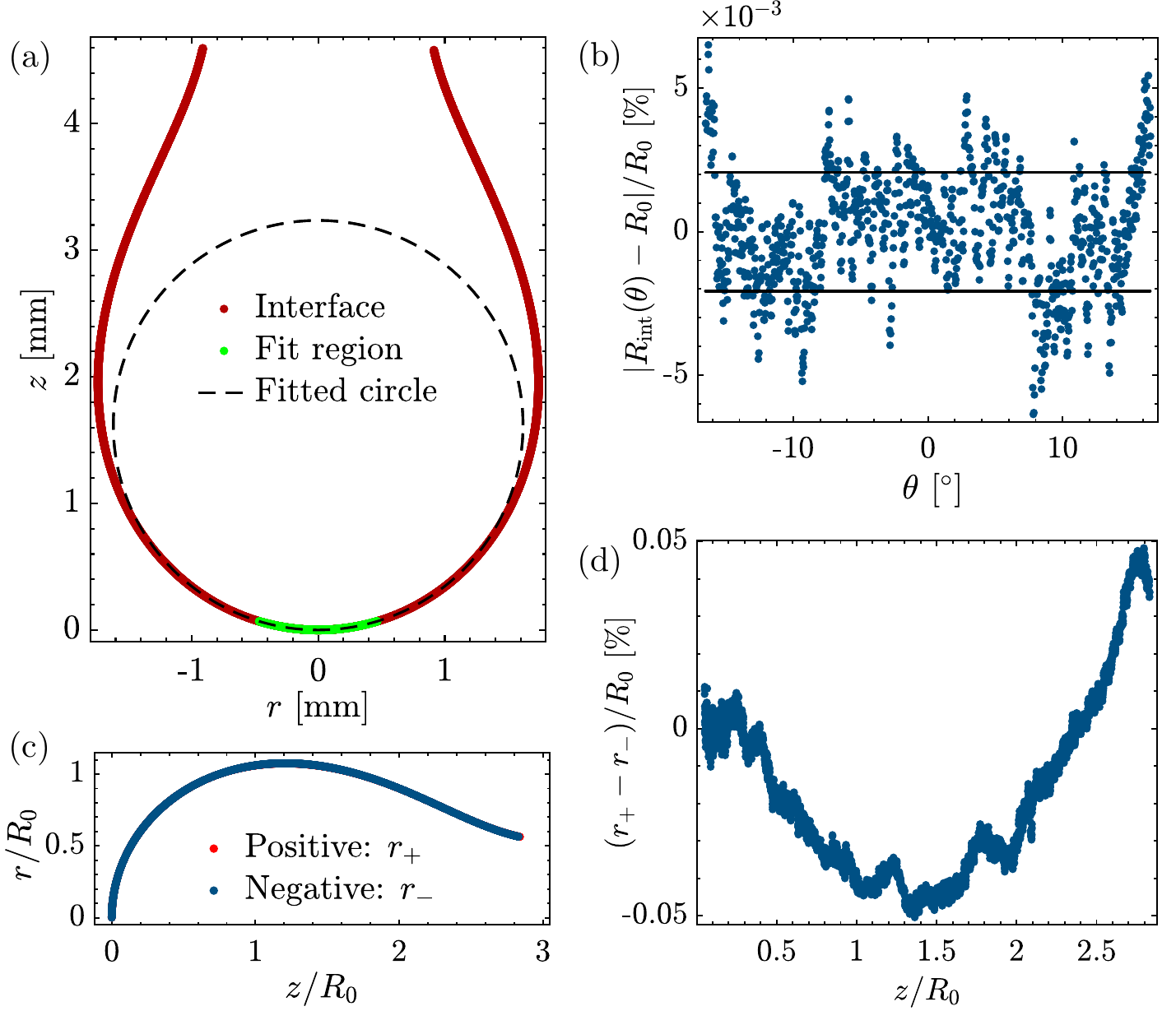}
    \caption{(a) Radius of curvature fit (circle) of the bottom part of the interface (highlighted with light green). (b) Normalised difference between the drop interface and the fitted circle. The horizontal lines indicate the uncertainty in the circle fit (standard deviation). (c) Both the positive half ($r>0$) and the negative half ($r>0$) of the drop interface are shown as a function of $z$. (d) The difference between both halves of the drop as a function of $z$.}
    \label{fig:exp_R0_fit}
\end{figure}

The detected drop edges are first converted from px to mm using the resolution fit (Fig. \ref{fig:resolution}) and rotated by the camera angle fit (Fig. \ref{fig:angle}). Then the bottom 1.5 $\%$ of the drop interface is selected and a circle is fitted to determine $R_0$, as shown in Fig. \ref{fig:exp_R0_fit}(a). The value of $1.5\%$ was chosen to be sufficiently large to obtain a accurate fit, but sufficiently small such that the effect of gravity was minimal. Fig. \ref{fig:exp_R0_fit}(b) shows the difference between the fitted circle and the detected interface. The uncertainty in the fit of the radius of curvature was calculated as the standard deviation of the difference between the fitted circle and the detected interface. Based on the circle fit, coordinates of the bottom of the drop were determined and the drop interface was split in an positive part (for $r > 0$) and a negative part (for $r < 0$), which are both shown in Fig. \ref{fig:exp_R0_fit}(c). The difference between both halves of the drop is shown in Fig. \ref{fig:exp_R0_fit}(d).

\begin{figure}[t!]
    \centering
    \includegraphics[width=0.75\textwidth]{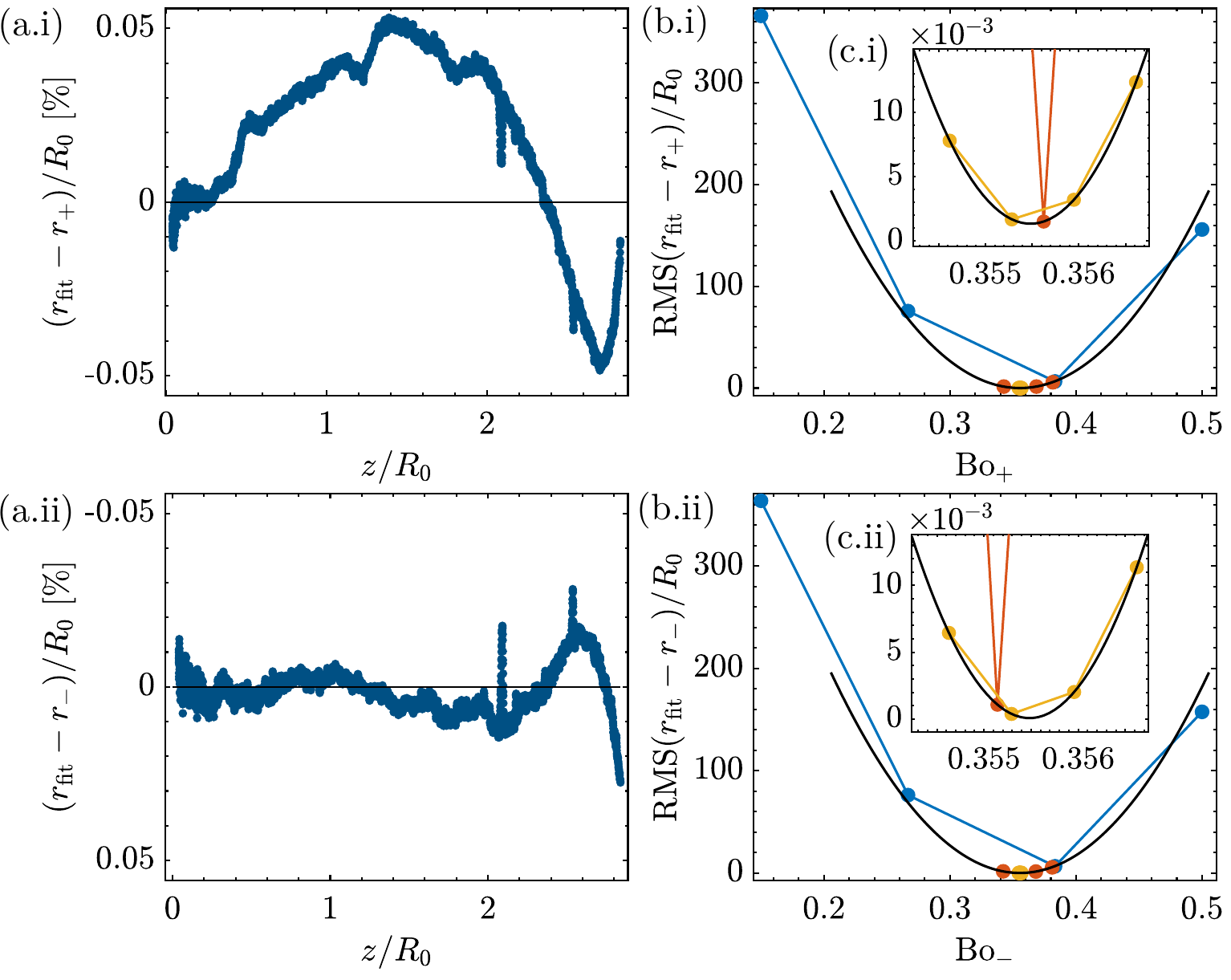}
    \caption{(a.i) and (a.ii) show the normalised difference between the YL-fit and the drop interface for the positive and negative side respectively as a function of $z$. (b.i) and (b.ii) show the root mean square of the difference between the YL-fit and the drop interface as a function of the Bo number. Different iterations are shown, each spanning a smaller range of  Bo numbers. The residues show a parabolic dependence on Bo, therefore a quadratic fit is used to determine the Bo number. Each inset, (c.i) and (c.ii) show a close up of the minimum. The parabola shown is a fit of the last iteration only (i.e. the yellow dots.)}
    \label{fig:exp_Bo_fit}
\end{figure}

Next, the non-dimensional coordinates of the interface $r/R_0$ as a function of $z/R_0$ are compared to the solution to the Young-Laplace equation (\ref{eq:fit_YL_eq}) for different Bo numbers. The Bo number is fitted using a least square method as is shown in Fig. \ref{fig:exp_Bo_fit}. To optimise for computational time, the best Bo number was determined using iterations of smaller intervals of Bo numbers. The first iteration encompassed all possible Bo numbers for this setup. Next, a quadratic fit is made of the residues (root mean square, RMS, of the difference between the YL-fit and the drop interface) and the next interval of Bo numbers is centered around this minimum, see Fig. \ref{fig:exp_Bo_fit}(b,c). In just three iterations in combination with the quadratic fit of the residues the Bo number could be with less than $4\cdot10^{-7}$ uncertainty, which is much smaller than any of the experimental errors in Bo. The difference between the YL-fit and the drop interface for the final Bo number is shown in Fig. \ref{fig:exp_Bo_fit}(a). The maximum difference is 0.05 $\%$, which is less than 1 $\tcmu m$ or less than 1 pixel as show in Fig. \ref{fig:drop_and_fit}.

\FloatBarrier
\clearpage
\section{All pendant drop measurements}
\begin{figure}[tp]
    \centering
    \includegraphics[width=1\textwidth]{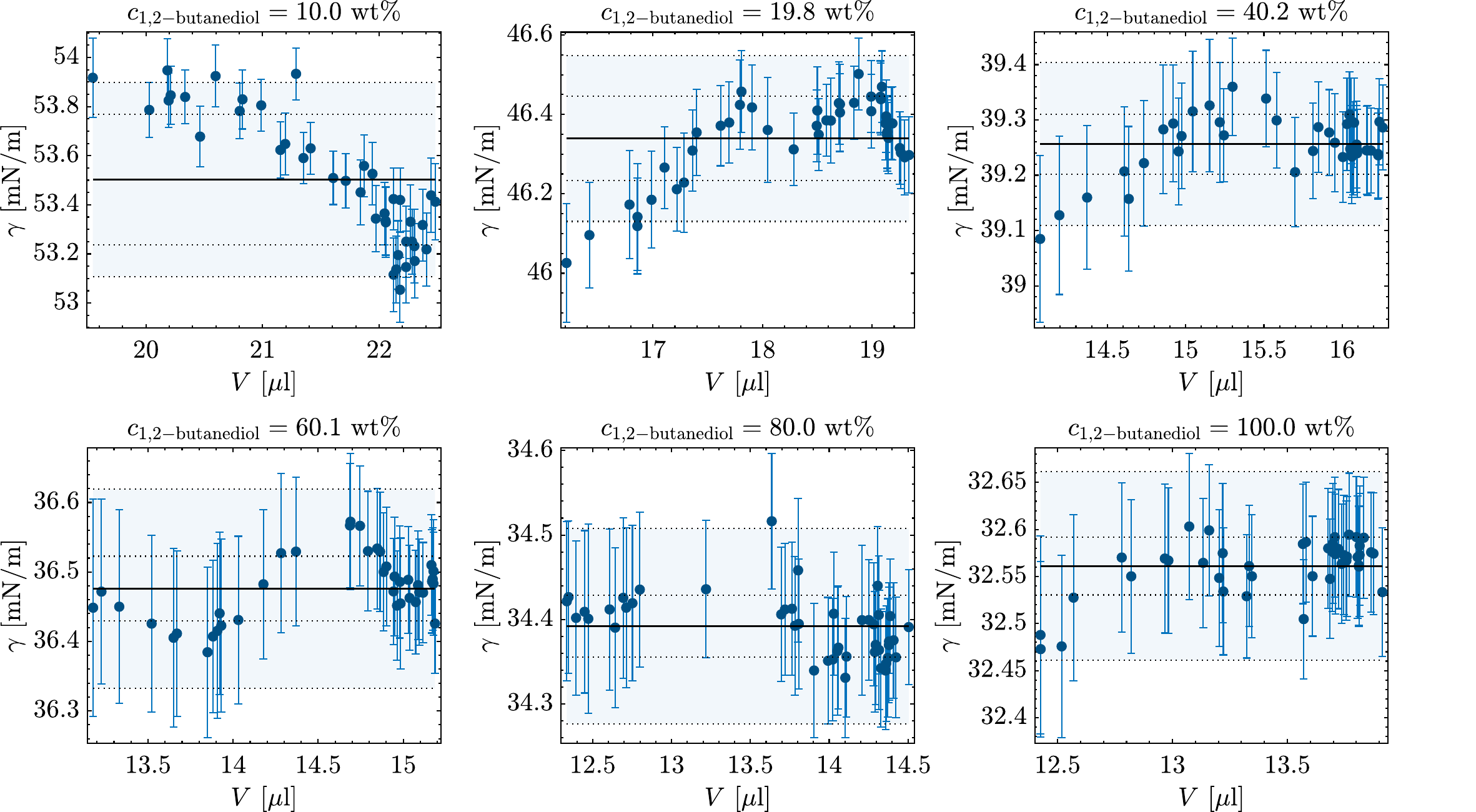}
    \caption{All individual measurements for all water/1,2-butanediol solutions.}
    \label{fig:gamma_w_bd_all}
\end{figure}

\begin{figure}[tp]
    \centering
    \includegraphics[width=1\textwidth]{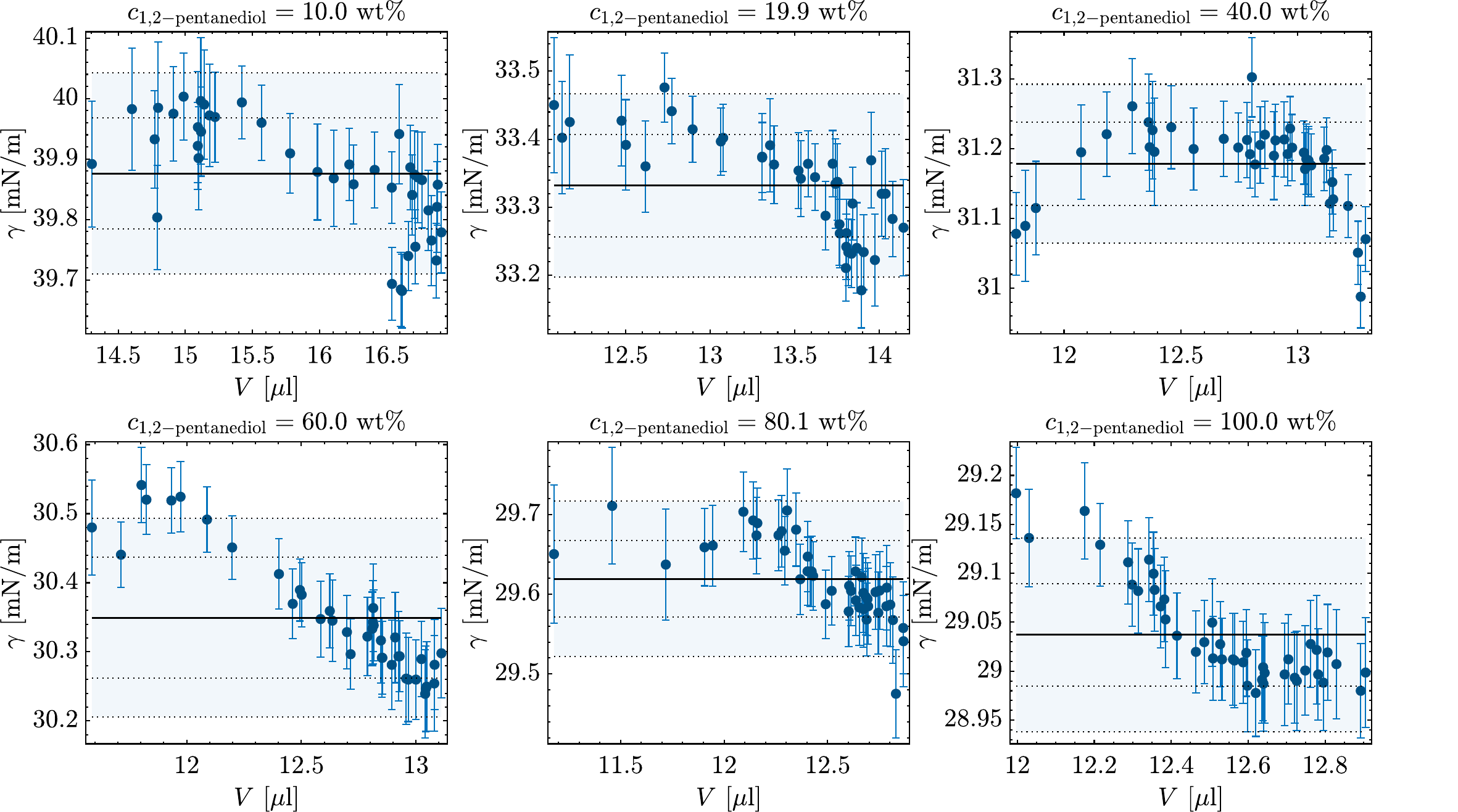}
    \caption{All individual measurements for all water/1,2-pentanediol solutions.}
    \label{fig:gamma_w_pd_all}
\end{figure}

\begin{figure}[tp]
    \centering
    \includegraphics[width=1\textwidth]{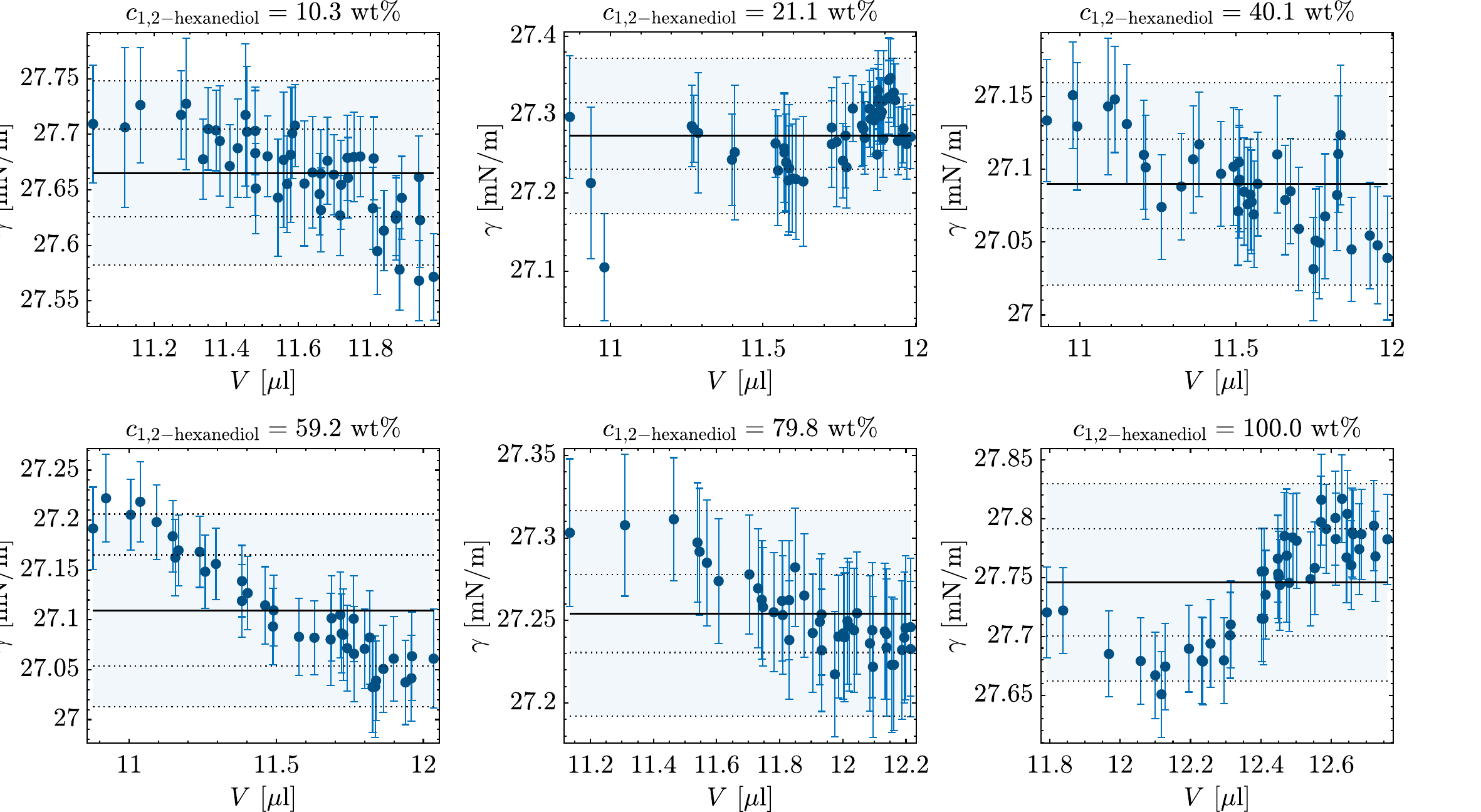}
    \caption{All individual measurements for all water/1,2-hexanediol solutions.}
    \label{fig:gamma_w_hd_all}
\end{figure}

\begin{figure}[tp]
    \centering
    \includegraphics[width=1\textwidth]{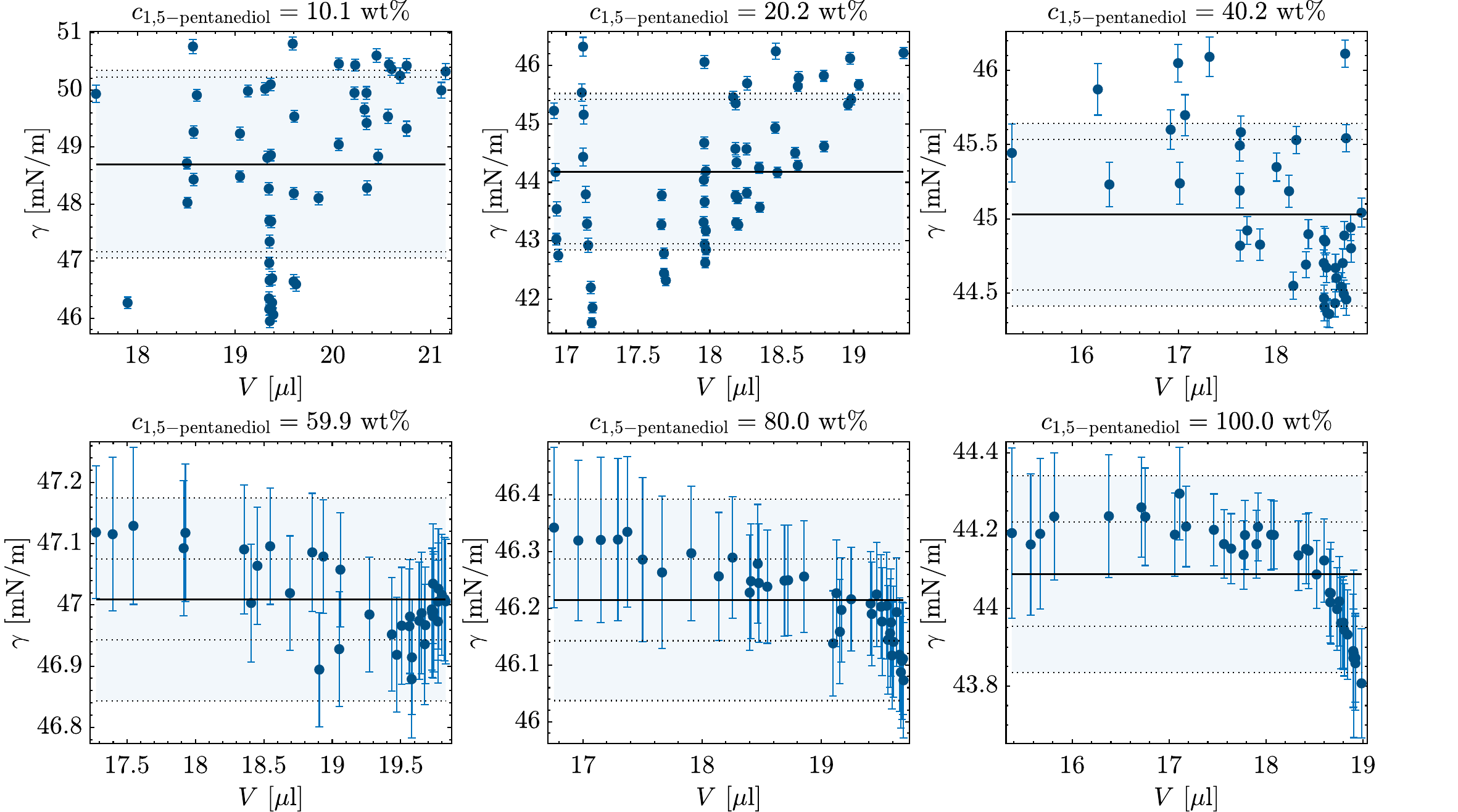}
    \caption{All individual measurements for all water/1,5-pentanediol solutions.}
    \label{fig:gamma_w_15_all}
\end{figure}

\begin{figure}[H]
    \centering
    \includegraphics[width=1\textwidth]{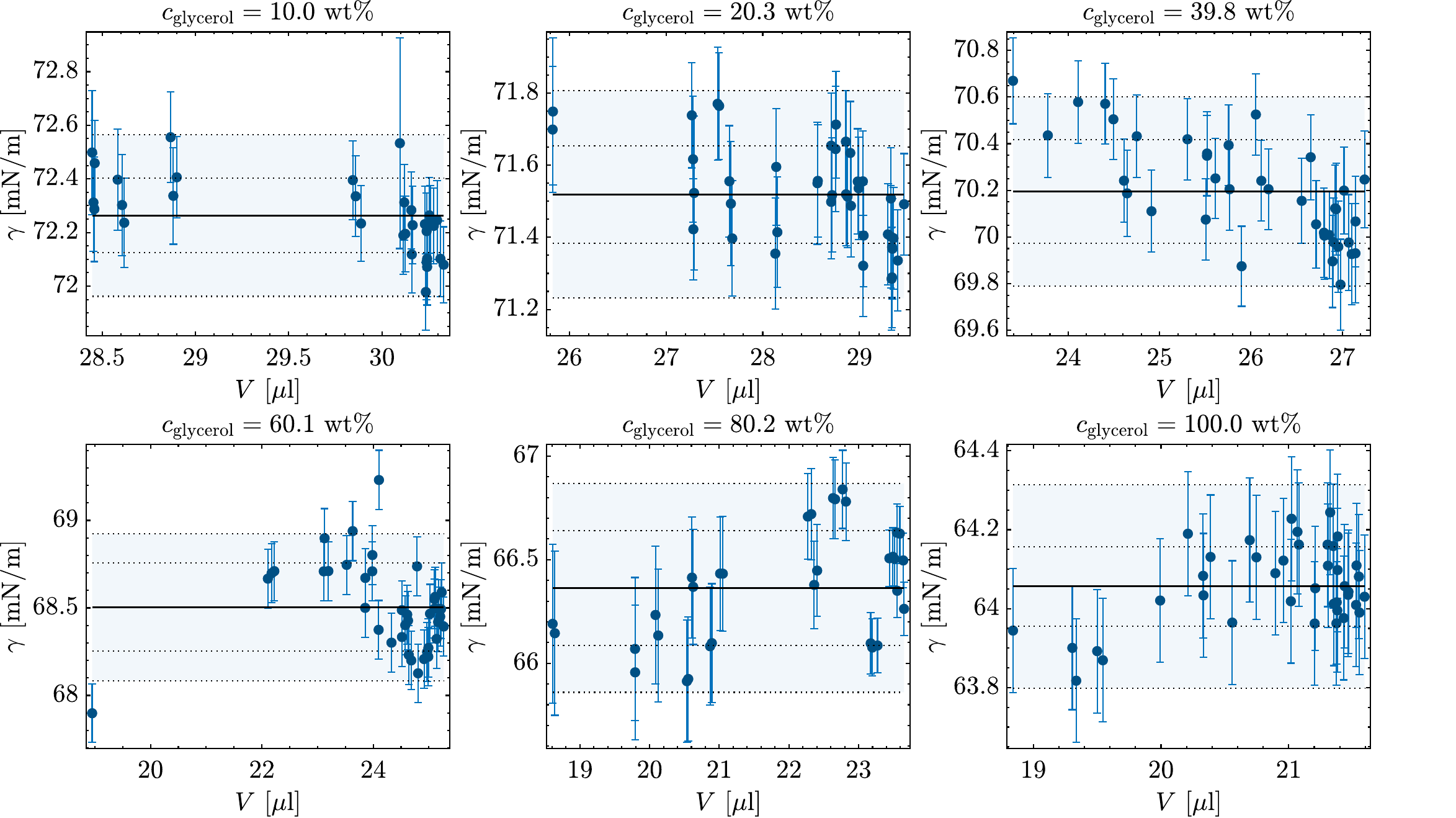}
    \caption{All individual measurements for all water/glycerol solutions.}
    \label{fig:gamma_w_gl_all}
\end{figure}

Table \ref{tab:gamma_all} shows all the densities and surface tensions measured using the pendant drop method. The values of the surface tension are average values of many repetitions of the pendant drop measurement for each solution. Figs. \ref{fig:gamma_w_bd_all} through \ref{fig:gamma_w_gl_all} show all individual measurements of the surface tension for each solution as a function of the drop volume. For each individual measurement the uncertainties is shown as an error bar. The horizontal black line is the average surface tension of all measurements (also shown in Tab. \ref{tab:gamma_all}). The standard deviation of all individual measurements (internal uncertainty) is indicated by the inner discontinuous line. The total uncertainty for each solution (also shown in Tab. \ref{tab:gamma_all}) is the combined internal uncertainty of all measurements and average individual uncertainty for each measurements for the surface tension. In Figs. \ref{fig:gamma_w_bd_all} through \ref{fig:gamma_w_gl_all} this is shown as the outer discontinuous line and the shaded region.

\begin{table}[H]
    \caption{\label{tab:gamma_all}
    Concentration, measured density, and average measured surface tension for all mixtures at $T = 21.0 \pm 0.2 \mathrm{^\circ C}$.
    }
    \begin{ruledtabular}
    \begin{tabular}{r r r r r r r}
        \multicolumn{3}{c}{\textbf{water/1,2-butanediol}} & $\quad$ & \multicolumn{3}{c}{\textbf{water/1,2-pentanediol}}\\
        $c$ [wt\%] & $\rho$ [kg/m\textsuperscript{3}] & $\gamma$ [mN/m]& $\quad$  & $c$ [wt\%] & $\rho$ [kg/m\textsuperscript{3}] & $\gamma$ [mN/m] \\
    \colrule
        $ 0.00    \pm  0.00 $  &  $ 998.0   \pm  0.5 $  &  $ 72.62  \pm  0.35 $  & $\quad$  & $ 0.00    \pm  0.00 $  &  $ 998.0   \pm  0.5 $  &  $ 72.62  \pm  0.35 $  \\
        $ 9.97    \pm  0.01 $  &  $ 1003.2  \pm  0.5 $  &  $ 53.50  \pm  0.40 $  & $\quad$  & $ 9.97    \pm  0.01 $  &  $ 1000.4  \pm  0.5 $  &  $ 39.88  \pm  0.17 $  \\
        $ 19.79   \pm  0.01 $  &  $ 1009.3  \pm  0.5 $  &  $ 46.34  \pm  0.21 $  & $\quad$  & $ 19.92   \pm  0.01 $  &  $ 1003.1  \pm  0.5 $  &  $ 33.33  \pm  0.13 $  \\
        $ 40.17   \pm  0.01 $  &  $ 1018.6  \pm  0.5 $  &  $ 39.26  \pm  0.15 $  & $\quad$  & $ 40.02   \pm  0.01 $  &  $ 1000.3  \pm  0.5 $  &  $ 31.18  \pm  0.11 $  \\
        $ 60.05   \pm  0.09 $  &  $ 1018.8  \pm  0.5 $  &  $ 36.48  \pm  0.14 $  & $\quad$  & $ 60.00   \pm  0.01 $  &  $ 993.8   \pm  0.5 $  &  $ 30.35  \pm  0.14 $  \\
        $ 79.99   \pm  0.01 $  &  $ 1013.1  \pm  0.5 $  &  $ 34.39  \pm  0.12 $  & $\quad$  & $ 80.12   \pm  0.01 $  &  $ 984.1   \pm  0.5 $  &  $ 29.62  \pm  0.10 $  \\
        $ 100.00  \pm  0.00 $  &  $ 1001.9  \pm  0.5 $  &  $ 32.56  \pm  0.10 $  & $\quad$  & $ 100.00  \pm  0.00 $  &  $ 970.7   \pm  0.5 $  &  $ 29.04  \pm  0.10 $  \\
    \colrule
    \colrule
        \multicolumn{3}{c}{\textbf{water/1,2-hexanediol}} & $\quad$ & \multicolumn{3}{c}{\textbf{water/1,5-pentanediol}}\\
        $c$ [wt\%] & $\rho$ [kg/m\textsuperscript{3}] & $\gamma$ [mN/m]& $\quad$  & $c$ [wt\%] & $\rho$ [kg/m\textsuperscript{3}] & $\gamma$ [mN/m] \\
    \colrule
        $ 0.00    \pm  0.00 $  &  $ 998.0   \pm  0.5 $  &  $ 72.62  \pm  0.35 $  & $\quad$  & $ 0.00    \pm  0.00 $  &  $ 998.0   \pm  0.5 $  &  $ 72.62  \pm  0.35 $  \\
        $ 10.27   \pm  0.01 $  &  $ 997.6   \pm  0.5 $  &  $ 27.67  \pm  0.08 $  & $\quad$  & $ 10.11   \pm  0.01 $  &  $ 998.4   \pm  0.5 $  &  $ 48.70  \pm  1.64 $  \\
        $ 21.12   \pm  0.01 $  &  $ 993.2   \pm  0.5 $  &  $ 27.27  \pm  0.10 $  & $\quad$  & $ 20.18   \pm  0.00 $  &  $ 1000.7  \pm  0.5 $  &  $ 44.18  \pm  1.34 $  \\
        $ 40.05   \pm  0.00 $  &  $ 984.8   \pm  0.5 $  &  $ 27.09  \pm  0.07 $  & $\quad$  & $ 40.17   \pm  0.00 $  &  $ 1004.1  \pm  0.5 $  &  $ 45.03  \pm  0.61 $  \\
        $ 59.18   \pm  0.03 $  &  $ 975.8   \pm  0.5 $  &  $ 27.11  \pm  0.10 $  & $\quad$  & $ 59.91   \pm  0.01 $  &  $ 1004.5  \pm  0.5 $  &  $ 47.01  \pm  0.17 $  \\
        $ 79.83   \pm  0.01 $  &  $ 966.5   \pm  0.5 $  &  $ 27.25  \pm  0.06 $  & $\quad$  & $ 79.95   \pm  0.00 $  &  $ 1000.9  \pm  0.5 $  &  $ 46.21  \pm  0.18 $  \\
        $ 100.00  \pm  0.00 $  &  $ 951.1   \pm  0.5 $  &  $ 27.75  \pm  0.08 $  & $\quad$  & $ 100.00  \pm  0.00 $  &  $ 989.5   \pm  0.5 $  &  $ 44.09  \pm  0.25 $  \\
    \colrule
    \colrule
        \multicolumn{3}{c}{\textbf{water/gycerol}}                               & $\quad$  & \multicolumn{3}{c}{ }\\
        $c$ [wt\%]  & $\rho$ [kg/m\textsuperscript{3}] & $\gamma$ [mN/m]         & $\quad$  &  &  & \\
    \colrule
        $ 0.00    \pm  0.00 $  &  $ 998.0   \pm  0.5 $  &  $ 72.62  \pm  0.35 $  & $\quad$  & & & \\
        $ 10.04   \pm  0.15 $  &  $ 1020.7  \pm  0.5 $  &  $ 72.26  \pm  0.30 $  & $\quad$  & & & \\
        $ 20.31   \pm  0.03 $  &  $ 1046.0  \pm  0.5 $  &  $ 71.52  \pm  0.29 $  & $\quad$  & & & \\
        $ 39.79   \pm  0.01 $  &  $ 1097.5  \pm  0.5 $  &  $ 70.20  \pm  0.41 $  & $\quad$  & & & \\
        $ 60.10   \pm  0.01 $  &  $ 1152.7  \pm  0.5 $  &  $ 68.50  \pm  0.42 $  & $\quad$  & & & \\
        $ 80.18   \pm  0.01 $  &  $ 1205.4  \pm  0.5 $  &  $ 66.36  \pm  0.50 $  & $\quad$  & & & \\
        $ 100.00  \pm  0.00 $  &  $ 1254.6  \pm  0.5 $  &  $ 64.06  \pm  0.26 $  & $\quad$  & & & \\

    \end{tabular}
    \end{ruledtabular}
\end{table}

\FloatBarrier


\bibliography{main.bib}

\end{document}